\documentclass[12pt,preprint]{aastex}

\shorttitle{Infall and Outflow in Sgr B2}

\shortauthors{Qin et al.}

\begin{document}




\title{Infall and Outflow of Molecular Gas in Sgr B2}







\author{Sheng-Li Qin \altaffilmark{1,2}, Jun-Hui Zhao \altaffilmark{1}, James M. Moran \altaffilmark{1},
Daniel P. Marrone \altaffilmark{1},\\
Nimesh A. Patel \altaffilmark{1}, Jun-Jie Wang\altaffilmark{2},
Sheng-Yuan Liu \altaffilmark{3}, Yi-Jehng Kuan \altaffilmark{3,4}}
\altaffiltext{1}{Harvard-Smithsonian Center for Astrophysics, 60
Garden Street, MS 42, Cambridge, MA 02138; sqin@cfa.harvard.edu.}
\altaffiltext{2}{National Astronomical Observatories, Chinese
Academy of Sciences, Beijing, 100012.} \altaffiltext{3}{Academia
Sinica Institute of Astronomy and Astrophysics, P.O.Box 23-141,
Taipei 106, Taiwan.} \altaffiltext{4}{Department of Earth
Sciences, National Taiwan Normal University, 88 Section 4,
Ting-Chou Road, Taipei 116, Taiwan.}






\begin{abstract}

Observations of two H$_2$CO ($3_{03}-2_{02}$ and $3_{21}-2_{20}$)
lines and continuum emission at 1.3 mm towards Sgr B2(N) and Sgr
B2(M) have been carried out with the SMA. The mosaic maps of Sgr
B2(N) and Sgr B2(M) in both continuum and lines show a complex
distribution of dust and molecular gas in both clumps and
filaments surrounding the compact star formation cores. We have
observed a decelerating outflow originated from the Sgr B2(M)
core, showing that both the red-shifted and blue-shifted outflow
components have a common terminal velocity. This terminal velocity
is 58$\pm$2 km s$^{-1}$. It provides an excellent method in
determination of the systematic velocity of the molecular cloud.
The SMA observations have also shown that a large fraction of
absorption against the two continuum cores is red-shifted with
respect to the systematic velocities of Sgr B2(N) and Sgr B2(M),
respectively, suggesting that the majority of the dense molecular
gas is flowing into the two major cores where massive stars have
been formed. We have solved the radiative transfer in a
multi-level system with LVG approximation. The observed H$_2$CO
line intensities and their ratios can be adequately fitted with
this model for the most of the gas components. However, the line
intensities between the higher energy level transition H$_2$CO
($3_{21}-2_{20}$) and the lower energy level transition H$_2$CO
($3_{03}-2_{02}$) is reversed in the red-shifted outflow region of
Sgr B2(M), suggesting the presence of inversion in population
between the ground levels in the two K ladders (K$_{-1}$= 0 and
2). The possibility of weak maser processes for the H$_2$CO
emission in Sgr B2(M) is discussed.

\end{abstract}














\section{Introduction}

The giant molecular cloud Sgr B2, located close to the Galactic
center ($\sim$44 arcmin from Sgr A*), is a well-known massive
star-forming region in our Galaxy. Sgr B2 consists of an extended
envelope, a hot ring and a few compact cores (e.g. Goicoechea,
Rodriguze-Fernandez \& Cernicharo 2004). The radio continuum and
recombination line observations of the compact HII regions suggest
that Sgr B2(N) and Sgr~B2~(M) are the two most active star forming
cores in this region (Gaume \& Claussen 1990; Gaume et al. 1995,
Mehringer et al. 1993; de Pree et al. 1995, 1996, 1998). Masers,
outflows and possible rotation of the two dense cores have been
revealed from observations of various molecular lines at
centimeter and millimeter wavelengths (Reid et al. 1988; Gaume \&
Claussen 1990; Martin-Pintado et al. 1990; Mehringer, Goss \&
Palmer 1994; Lis et al. 1993; Kuan \& Snyder 1996; Liu \& Snyder
1999). In addition, previous observations have shown evidence for
the two hot cores to be at different evolutionary stages and to
have different molecular abundances (e.g. Vogel et al. 1987; Lis
et al. 1993; Miao et al. 1995; Kuan, Mehringer \& Snyder 1996; Liu
\& Snyder 1999).

H$_2$CO pervades the interstellar medium and it has a simple
chemical reaction path which has been proven to be a useful probe
of physical conditions (e.g. Mangum \& Wootten 1993). The
H$_{2}$CO (1$_{10}-1_{11}$) transition at 6 cm was observed in
absorption against discrete continuum sources towards Sgr B2
complex with an angular resolution of
$\sim10^{\prime\prime}\times20^{\prime\prime}$, showing nearly the
same radial velocity pattern as that of the radio recombination
lines (Martin-Pintado et al. 1990; Mehringer, Palmer \& Goss
1995). These authors suggested that the H$_{2}$CO
(1$_{10}-1_{11}$) transition probably arises from the surrounding
gas with a relatively low mean H$_{2}$ density of $\sim 10^{4}$
cm$^{-3}$ (Martin-Pintado et al. 1990; Mehringer, Palmer \& Goss
1995).

The millimeter H$_{2}$CO lines are an excellent tracer of H$_{2}$
density $> 10^{5}$ cm$^{-3}$ (e.g., Mangum \& Wootten 1993). In
addition, H$_{2}$CO is a planar asymmetric top molecule with very
little asymmetry. The symmetry of the spin function of the
molecule leads to two transition classes: ortho-H$_{2}$CO levels
if the spin wavefunction is symmetric and para-H$_{2}$CO levels if
antisymmetric. Since para-H$_{2}$CO is 1-3 times less abundant
than ortho-H$_{2}$CO, observations of para-H$_{2}$CO have less
opacity effect (Kahane et al 1984; Mangum \& Wootten 1993). Hence,
para-H$_{2}$CO appears to be a better probe to determine the
physical conditions of the massive star formation regions.

The millimeter/submillimeter transitions of H$_{2}$CO gas require
relatively high excitation temperature and high H$_{2}$ density
compared to those in the centimeter wavebands. If the brightness
temperature of the continuum emission is higher than the
excitation temperature, the absorption against the continuum cores
can be observed in millimeter and sub-millimeter wavebands with
the high angular resolution of an interferometric array (such as
the Submillimeter Array,\footnote {The Submillimeter Array is a
joint project between the Smithsonian Astrophysical Observatory
and the Academia Sinica Institute of Astronomy and Astrophysics
and is funded by the Smithsonian Institution and the Academia
Sinica.} hereafter SMA). Taking advantage of the large bandwidth
coverage of the SMA, we have observed multiple H$_{2}$CO lines
towards Sgr B2 at 1.3 mm within a bandwidth of 2 GHz. Thus, with
the same telescope system and calibration procedure, the
uncertainties due to absolute flux density calibration among the
different line transitions can be mitigated by measuring the
line-intensity ratios, which are needed to determine physical
conditions, such as kinetic temperature and H$_{2}$ number
density, of the gas. In addition, the SMA is not sensitive to
extended larger scale emission ($\sim$ 50\arcsec). Thus, the SMA
observations are sensitive to the clumps of high density gas
rather than the extended diffuse components.

In this paper, we present the results from the SMA observations of
Sgr B2 at the H$_{2}$CO lines and continuum at 1.3 mm. The paper
is organized as follows: \S 2 discusses the observations and data
reduction. In \S 3 we present the data analysis and results. In \S
4, we present the kinematics in Sgr B2(M) by a model incorporating
a spherically symmetric inflow along with a decelerating outflow.
In \S 5 we model the physical properties of the H$_{2}$CO gas in
Sgr B2 using the large velocity gradient (LVG) approach. In \S 6,
we discuss the important results derived from our observations and
analysis. We summarize the results in \S7. We adopt a distance of
8 kpc to Sgr B2.

\section{Observations \& Data Reduction}

Observations towards Sgr~B2(N) and Sgr~B2(M) were carried out at
218 (lower sideband) and 228 GHz (upper sideband) with seven
antennas of the SMA in the compact-north array on August 1, 2005
for 8 hours. The projected baselines ranged from 5 to 50
k${\lambda}$. The weather was good during the observations with
$\tau$$\sim$0.09 at 225 GHz. The typical system temperature was
130 K. Sgr~B2(N) and Sgr~B2(M) were observed in separate fields
with 15 min observing time on each source interleaving 5 min on
the phase reference source Sgr A* ($<0.1$ mas, $\sim$3 Jy at 230
GHz, $\sim$45 arcmin from Sgr B2). Our two target fields were
centered at RA(J2000)=17$^{\rm h}$47$^{\rm m}$19$\rlap{.}^{\rm
s}$882,
DEC(J2000)=$-28^{\circ}$22$^{\prime}$18$\rlap{.}^{\prime\prime}$37
and RA(J2000)=17$^{\rm h}$47$^{\rm m}$20$\rlap{.}^{\rm s}$156,
DEC(J2000)=$-28^{\circ}$23$^{\prime}$03$\rlap{.}^{\prime\prime}$56,
for Sgr B2(N) and Sgr B2(M), respectively. In addition, Callisto
(4.1 Jy) and the QSO 3C454.3 (32 Jy) were also observed for the
flux-density and bandpass calibrations. The flux density was
estimated from Callisto with the assumption that its brightness
temperature was 120 K and its angular size was
1$\rlap{.}^{\prime\prime}$14. The three transitions H$_{2}$CO
(3$_{03}-2_{02}$), (3$_{22}-2_{21}$) and (3$_{21}-2_{20}$) were
positioned in the lower 2 GHz sideband (LSB). H$_{2}$CO
(3$_{22}-2_{21}$) emission appears to be blended with a transition
of CH$_{3}$OH and will be not used in the analysis in this paper.
The two unblended transitions, H$_{2}$CO, (3$_{03}-2_{02}$) and
(3$_{21}-2_{20}$), have upper level energies of 21.0 and 67.8 K
and rest frequencies of 218.2222 and 218.7601 GHz, respectively.
The spectral resolution of 0.8125 MHz corresponds to a velocity
resolution of 1.1 km~s$^{-1}$.

The data reduction was carried out in {\it Miriad} \footnote {The
data reduction procedures for SMA data are outlined in the web site
http://sma-www.cfa.harvard.edu/miriadWWW}. We recomputed the
Doppler velocity for each of the target sources since the on-line
Doppler tracking was only made on Sgr~B2(N). System temperature
corrections were applied. Antenna-based bandpass ripples were
corrected by applying the linear interpolation of the bandpass
solutions determined from Callisto and 3C454.3. There are
spectral-window-based offsets in amplitude and phase on some
baselines due to the correlator errors, which were also corrected.
The residual errors due to the bandpass shape of the individual
spectral windows were reduced to a level below 1\% of the
continuum level. In the antenna-based gain corrections, we chose
the visibilities of Sgr~A* in the UV range between 20 kilo
wavelengths and longer in order to eliminate the contamination
from the extended dust and HII emission. The gains determined from
the point source (emission from Sgr A*) were applied to the Sgr~B2
data.

The molecular lines were identified in the rest frame by use of
the JPL catalog, as done by others (Sutton et al. 1985 and
Nummelin et al. 1998). The H$_{2}$CO transitions and a few other
molecular lines have been unambiguously identified (see spectrum
in Fig. 1).

The continuum was subtracted with a linear fitting to the spectrum
of line-free channels in each baseline. For Sgr~B2(M), the
line-free channels can be easily selected from the spectrum in the
(u, v) domain (Fig. 1). However, for Sgr~B2(N), the molecular
spectral lines are crowded over each 2 GHz band and it is
difficult to choose the line-free channels. We developed a
procedure to select the line-free channels for the continuum
subtraction. First, after reducing of the spectral resolution to 1
km~s$^{-1}$, we made each of the channel maps (2600 channels over
the 2 GHz band) including both the line and continuum. From the
channel maps, we selected the channels without extended emission
($>$4$\sigma$). The continuum level was determined from these
apparently line-free channels. We selected 16 and 37 line-free
spectral windows for the fields of Sgr~B2(M) and Sgr~B2(N),
respectively. Then, using the task UVLIN in {\it Miriad}, we
determined the continuum level by fitting to the line-free
channels. The UVLIN gives two output (u, v) data sets, one for the
continuum and the other for continuum-free spectral line.

Self-calibration was performed to the continuum data using the
compact sources of the continuum cores for a few iterations in
order to eliminate residual errors. The gain solutions determined
from the continuum data were applied to the line data, from which
we constructed the spectral data cube. The preliminary images of
the continuum and lines were made using natural weighting. The
clean algorithm was applied to remove the effects of the
side-lobes. The synthesized beam sizes of the continuum and line
images were approximately 5$\rlap{.}^{\prime\prime}4\times
3\rlap{.}^{\prime\prime}2$ (PA=12.5$^{\circ}$). The mosaic maps of
the two fields, Sgr B2(N) and Sgr B2(M), were made using a simple
linear mosaic algorithm. The primary beam attenuation was also
corrected in the final line cubes and continuum image. The
statistical 1$\sigma$ rms noise of the continuum was 0.1
Jy~beam$^{-1}$. The 1$\sigma$ rms noise levels of the line images
were 0.18 and 0.16 Jy~beam$^{-1}$ per channel for the H$_{2}$CO
(3$_{03}-2_{02}$) and (3$_{21}-2_{20}$) lines, respectively. The
higher noise in the H$_{2}$CO (3$_{03}-2_{02}$) line images was
due to its strong line intensity and the limit of the dynamic
range in the clean process.

\section{Data Analysis \& Results}

\subsection{The Continuum at 1.3 mm}

Fig. 2 shows the mosaic continuum map of Sgr~B2(N) and Sgr~B2(M)
at 1.3 mm. The brightest components of Sgr~B2(N) (I$_{\rm
p}$=29.2$\pm$2.1 Jy~beam$^{-1}$) and Sgr~B2(M) (I$_{\rm
p}$=20.2$\pm$1.3 Jy~beam$^{-1}$), are associated with the massive
star forming cores K1-3 and F1-4 (Gaume \& Claussen 1990),
respectively. In addition to the emission from these cores, a few
nearby continuum clumps were detected, including the components K4
(I$_{\rm p}$=1.40$\pm$0.12 Jy~beam$^{-1}$), NE (I$_{\rm
p}$=1.20$\pm$0.14 Jy~beam$^{-1}$), NW (I$_{\rm p}$=1.00$\pm$0.11
Jy~beam$^{-1}$), ME (I$_{\rm p}$=0.58$\pm$0.15 Jy~beam$^{-1}$) and
MW (I$_{\rm p}$=1.70$\pm$0.18 Jy~beam$^{-1}$).

Gaussian fitting to the individual continuum components was
carried out. The individual emission clumps near the compact cores
or the K1-3 and F1-4 clusters were modeled as simple Gaussian
components. The two compact cores, K1-3 and F1-4, appeared to be
too complicated to be fitted with single Gaussian components. A
model consisting of a Gaussian and two unresolved compact
components was used to fit to the data of the core (K1-3). The
peaks of the Gaussian component and one of the point components
are consistent with K3 and K2 positions, respectively. Another
point component agrees (within $\sim$ 1$^{\prime\prime}$) with the
3.5 mm continuum source that is located south of K3 (see Fig.1(b)
of Liu \& Snyder 1999). Sgr B2(M) core (F1-4) is fitted well with
a Gaussian component and a point component (close to F3 within
$\sim$ 0$\rlap{.}^{\prime\prime}$5). The peak positions,
deconvolved angular sizes, peak intensities and total flux
densities of the continuum components are summarized in Table 1.

The components K4, MW and Z10.24 have been detected at radio and
millimeter wavelengths (Lis et al. 1993; Kuan \& Snyder 1994;
Gaume et al. 1995; Liu \& Snyder 1999). Z10.24 located in the
middle between Sgr B2(N) and Sgr B2(M) (see Fig. 2) shows a unique
filamentary structure at 1.3 mm continuum. The designation of
Z10.24 follows that used by Gaume et al. (1995) and  de Pree et
al. (1996) who detected the H66$\alpha$ line towards it. Z10.24
was marginally detected at 1.3 mm by Lis et al. (1993) with higher
angular resolution
(4$\rlap{.}^{\prime\prime}5\times3\rlap{.}^{\prime\prime}7$) and
poorer sensitivity. The SMA observations show an elongated
structure in Z10.24 with a peak intensity of 2.02 $\pm$ 0.16 Jy
beam$^{-1}$ (12 $\sigma$). The core of Z10.24 is unresolved in the
sub-arcsec resolution images at both 1.3 cm and 3.5 mm (Gaume et
al. 1995; Liu \& Snyder 1999). The 1.3 mm continuum image and the
detected vibrational HC$_{3}$N emission (de Vicente et al. 2000)
suggest that the 1.3 mm continuum of Z10.24 is dominated by the
dust emission and Z10.24 is likely to be a younger massive star
formation region.

There have been no detections of the components NW, NE and ME in
the previous observations at longer wavelengths. These three
components are possibly the dust emission from sub-cores at a
relatively early stage of star formation. The detections need to
be verified with further observations at shorter wavelengths and
at higher angular resolutions.

\subsection{H$_{2}$CO Lines}

The continuum-free channel maps in both H$_{2}$CO
(3$_{03}-2_{02}$) and H$_{2}$CO (3$_{21}-2_{20}$) lines were
constructed in the velocity range from 8 to 151 km~s$^{-1}$ at
intervals of 1 km~s$^{-1}$. The channel maps of the H$_{2}$CO
transitions are complicated, containing several kinematical
features in either emission or absorption. Those emission and
absorption components are separated well in our higher spectral
resolution maps but some of them (in the continuum core regions)
are overlapped. In the moment analysis, the negative intensity
value from the absorption and the positive value from the emission
may cancel each other in the overlapping regions and thus the
resultant moment maps might not reflect the true gas distribution.
Hence, the emission and absorption need to be handled separately.

\subsubsection{Line Emission}

Fig. 3 is the integrated line emission images constructed from the
channel maps for the two transitions, H$_{2}$CO ($3_{21}-2_{20}$)
and ($3_{03}-2_{02}$), respectively. The moment 0 images were made
with a 4 $\sigma$ cutoff in each channel maps. The less
significant emission and the absorption are excluded in the moment
analysis. Most of emission is distributed around the two cores of
Sgr~B2(N) and Sgr~B2(M). Clearly, the distribution of the
H$_{2}$CO emission is not spherically symmetric with respect to
each of these cores. In the Sgr~B2(N) region, in addition to the
gas concentration at the core, gas clumps located north-east and
south-west of the core are observed in both H$_{2}$CO transitions.
We note that in Sgr~B2(N), the morphology of H$_{2}$CO
($3_{21}-2_{20}$) emission, the higher transition gas, is similar
to that of the emission from the lower transition gas H$_{2}$CO
($3_{03}-2_{02}$). Hereafter we refer H$_{2}$CO ($3_{21}-2_{20}$)
as the higher transition and H$_{2}$CO ($3_{03}-2_{02}$) as the
lower transition. In the Sgr~B2(M) region, a strong emission
component elongated in northwest-southeast direction is observed
in both H$_{2}$CO transitions. An arch structure (M1~
30$^{\prime\prime}$ long and 10$^{\prime\prime}$ wide),
8$^{\prime\prime}$ northwest of the Sgr~B2(M) core appears in both
H$_{2}$CO transitions. The major difference in the emission
distribution from the two transitions occurs in the outflow (Lis
et al. 1993) region located south-east of the Sgr~B2(M) core. A
significant emission ``tongue'' (M5)
(15$^{\prime\prime}\times7^{\prime\prime}$) was detected from the
higher H$_{2}$CO transition gas while no significant detection was
made of the lower H$_{2}$CO transition gas.

\subsubsection{Systematic velocities}

Based on a line survey at 340 GHz from single dish observations,
Sutton et al. (1991) obtained mean systematic velocities 65 and 61
km~s$^{-1}$ of Sgr B2(N) and Sgr B2(M). Their observations showed
significant velocity variations among the different species. The
differences are mostly caused by the chemical differences of the
molecules. The different species sample different physical
environments. The high angular resolution observations of
H66$\alpha$ (de Pree et al. 1995; 1996) showed that the mean
systematic velocities are 69.8 and 65.3 km~s$^{-1}$ for the Sgr
B2(M)-F and Sgr B2(N)-K clusters, respectively. The relatively
higher mean velocity of 69.8 km~s$^{-1}$ in Sgr B2(M) is likely
caused by the high velocity motion of the ionized gas of the UCHII
regions with respect to the centroid of the system. In section 4,
we will show that the systematic velocity determined from the
terminal velocity of the outflow in Sgr~B2(M) is 58 km s$^{-1}$
(see also Appendix A). In the rest of the paper, we adopt
systematic velocities of 58 km s$^{-1}$ and 65 km s$^{-1}$ for
Sgr~B2(M) and Sgr~B2(N), respectively.

\subsubsection{Absorption towards the cores}

The strong continuum cores are excellent probes of absorption by
the cold gas that resides in front of them. Absorption is observed
towards both Sgr B2(N) and Sgr B2(M) continuum cores (see the
top-left and bottom-left panels of Fig. 3). Multiple Gaussian line
components were fitted to the spectra for each of the two
transitions in both cores. The parameters of these fits are listed
in Table 2. The systematic velocities are marked with the vertical
lines in Fig. 3. The majority of the absorbing gas in Sgr~B2(N)
and Sgr~B2(M) is red-shifted with respect to the systematic
velocities 58 and 65 km~s$^{-1}$, respectively. The red-shifted
absorption gas provides evidence for the existence of gas
accreting onto the two cores.

If the absorbing gas covers the continuum source completely and
the emission from the gas is insignificant, the line intensity is
${\rm I}_{\rm L} \approx {\rm I}_{\rm C} e^{-\tau_{\rm L}}$. The
optical depth (${\tau}_{\rm L}$) can be derived from the formula

\begin{equation}
\tau_{\rm L}=-ln(\frac{{\rm I}_{\rm L}}{{\rm I}_{\rm
C}})=-ln(1+\frac {\Delta {\rm I}_{\rm L}}{{\rm I}_{\rm C}}),
\end{equation}

\noindent where $\Delta {\rm I}_{\rm L} = {\rm I}_{\rm L}-{\rm
I}_{\rm C}$ is the observed line intensity and the ${\rm I}_{\rm
C}$ is the observed continuum intensity. The errors in $\tau_{\rm
L}$ based on the fractional errors $\sigma_{{\rm I}_{\rm L}}/{\rm
I}_{\rm L}$ of each line channel and $\sigma_{{\rm I}_{\rm
C}}/{\rm I}_{\rm C}$ if the line is optically thin ($\displaystyle
-\frac{\Delta {\rm I}_{\rm L}}{{\rm I}_{\rm C}}< 1-e^{-1} \approx 0.6$)
are given by

\begin{equation}
       \sigma_{ \tau}\approx \sqrt{(\sigma_{
       {\rm I}_{\rm L}}/{\rm I}_{\rm L})^{2}+(\sigma_{{\rm I}_{\rm C}}/{\rm I}_{\rm
       C})^{2}}.
\end{equation}

\noindent In the optically thick case, the channels in the line
center are saturated and the line-to-continuum ratio only gives a
lower limit to the optical depth of the line. For example if $n$
spectral channels are saturated by the absorption gas component at
a velocity, we have ${\rm I}_{\rm L}< 3\sigma_{ {\rm I}_{\rm
L}}/\sqrt{n}$ in 3 ${\sigma}$. Substituting this formula into the
equation (1), the lower limit of the optical depth is

\begin{equation}
\tau_{\rm L}>-ln(\frac {3\sigma_{{\rm I}_{\rm L}}}{\sqrt{n}{\rm
I}_{\rm C}}).
\end{equation}

With the Rayleigh-Jeans approximation, the value of 1
Jy~beam$^{-1}$ in our SMA observations is equivalent to 1.5 K. The
observed peak continuum intensities of the Sgr~B2(N) and Sgr~B2(M)
cores are 29.2 and 20.2 Jy~beam$^{-1}$, which correspond to
brightness temperature of 44 and 30 K in our observations,
respectively. The solution of radiation transfer function in terms
of the observed brightness temperature of the line ($\Delta {\rm
T}_{\rm L}^{\rm obs}$) is given by

\begin{equation}
    \Delta {\rm T}_{\rm L}^{\rm obs}=( f_{\rm L}{\rm T}_{\rm ex}-f_{o}{\rm T}^{\rm obs}_{\rm C})
(1-e^{-{\tau_{\rm L}}}),
\end{equation}

\noindent where ${\rm T}_{\rm ex}$ is the excitation temperature
of the molecular line and ${\rm T}_{\rm C}^{\rm obs} = f_{\rm C}
{\rm T}_{\rm C}$ is the observed brightness temperature of the
continuum emission with the true brightness temperature of ${\rm
T}_{\rm C}$; $\tau_{\rm L}$ is the optical depth of the molecular
cloud; for given solid angles of the molecular cloud ($\Omega_{\rm
L}$), continuum source ($\Omega_{\rm C}$) and the telescope beam
($\Omega_{\rm B}$), $f_{\rm L}=\Omega_{\rm L}/(\Omega_{\rm
L}+\Omega_{\rm B})$ and $f_{\rm C}=\Omega_{\rm C}/(\Omega_{\rm
C}+\Omega_{\rm B})$ are the beam filling factors of the line and
continuum, respectively, if both the source and the telescope beam
are in Gaussian shape; $f_0$ denotes the fraction of the continuum
source covered by the molecular cloud.

If the molecular cloud is in front of the larger continuum source
($f_0 \le {\displaystyle f_{\rm L} \over \displaystyle f_{\rm
C}}$), the observed brightness temperature of the line (${\rm
T}_{\rm L}^{\rm obs}$) becomes:

\begin{equation}
       \Delta {\rm T}_{\rm L}^{\rm obs}=( f_{\rm L}{\rm T}_{\rm ex}-{f_{\rm L}\over f_{\rm C}}{\rm T}^{\rm obs}_{\rm
       C})(1-e^{-{\tau_{\rm L}}}).
\end{equation}

\noindent If the molecular clouds cover the continuum cores with
the same beam filling factor ($f_{\rm L}=f_{\rm C}$), the upper
limits of the excitation temperatures H$_{2}$CO absorbing gas
($\Delta {\rm T}_{\rm L}^{\rm obs}< 0$) are imposed by the
observed continuum brightness temperatures of the continuum cores,
i.e. 44/$f_{\rm L}$ and 31/$f_{\rm L}$ K, for the Sgr~B2(N) and
Sgr~B2(M) cores. For a given filling factor $f_{\rm L}=0.3$, the
excitation temperatures of H$_{2}$CO would be less than 150 and
103 K for Sgr~B2(N) and Sgr~B2(M), respectively.

\subsubsection{Individual components}

Fig.~4 shows the spectra of the H$_{2}$CO lines (averaged over one
beam) made from the image cube for the rest of components. Each of
spectral panels includes both the lower (3$_{03}-2_{02}$, the
green profile) and the higher (3$_{21}-2_{20}$, the red profile)
transition lines. Gaussian fits to the H$_{2}$CO spectra were
carried out for both absorption and emission components. The
emission and absorption probably come from the different regions
along the line of sight. The angular resolution of our
observations appears to be inadequate to distinguish the discrete
components in the cores. However, our high spectral resolution is
adequate to separate the emission from the absorption in the
Gaussian fits. The parameters derived from the Gaussian fits to
the two H$_{2}$CO transitions are summarized in Table 2 including
the central line velocity (V$_{\rm LSR}$), the full width of half
maximum ($\Delta$V) and peak intensity (I$_{\rm p}$).

The M1, M4 and M5 are located along the major axis of the bipolar
outflow originated from the F-cluster (Lis et al. 1993). M1 is on
the blue-shifted side of the outflow. Both the high- and
low-transition spectra can be fitted with two Gaussian components
at 52 km s$^{-1}$ in emission and 64 km s$^{-1}$ in absorption.
The weak red-shifted absorption with respect to the systematic
velocity suggests that a relatively cold gas component in front of
the continuum source are moving towards it. The strong
blue-shifted emission with respect to the systematic velocity is
the highly-excited gas emission in the outflow (in front of the
continuum source) likely mixed with the infall gas emission
(behind the continuum source).

Sgr~B2(M)-M4 is located close to the HII regions, {\it i.e.}, the
F1-4 cluster. The higher transition spectrum can be fitted with
two Gaussian components in emission at 66 and 70 km s$^{-1}$, both
of which are red-shifted with respect to the systematic velocity
58 km s$^{-1}$. The lower transition spectrum can be fitted with
five Gaussian components at velocity 51, 65, 70, 75 km s$^{-1}$ in
absorption and 100 km s$^{-1}$ in emission. The morphology of this
region is a complex.

A possible model to interpret the spectral characteristics of M4
is considered here. If we assume a non-LTE condition for the gas
and that both higher and lower transition gas comes from the same
gas clump located in front of the continuum core, the excitation
temperature of the higher transition gas is larger than the
brightness temperature of the continuum, while the excitation
temperature of the lower transition gas is less than the
brightness temperature of the continuum. The red-shifted
absorption suggests that the gas flows towards the continuum core.
The nature of absorption and emission at M4 suggests that the
process of excitation of the infall molecular cloud is
complicated. The radiative excitation by the strong FIR radiation
field might play an important role in the region near the core
since the collision alone can not produce the observed line ratio
or the inverse population between the higher transition in
K$_{-1}$=2 and the lower transition in K$_{-1}$=0 based on the LVG
fitting (see Appendix B).

The inversion in population between the two lowest K ladders is
observed best in the red-shifted  outflow component. At M5, the
spectrum of the higher transition gas shows a very significant
line emission (7-8 $\sigma$), fitted to a Gaussian at 66 km
s$^{-1}$ with a line width of 16 km s$^{-1}$, while less
significant emission (1-2 $\sigma$) of the lower transition gas is
shown at the same position. The highly reversed line ratio
${\displaystyle {\rm I}(3_{21}-2_{20})\over \displaystyle {\rm
I}(3_{03}-2_{02})}\sim 6$ suggests that local thermodynamic
equilibrium (LTE) is not valid at this location and a weak maser
process is active in the outflow region.

The gas components M2 and M3 and MW show significant line emission
($>10$ $\sigma$) from the lower transition gas while the higher
transition  emission is relatively weak. The line ratio
${\displaystyle {\rm I}(3_{03}-2_{02}) \over \displaystyle {\rm
I}(3_{21}-2_{20}) }$ in those isolated component varies in the
range between 1.5 to 5.2.

H$_{2}$CO maser at 6 cm (Mehringer, Goss \& Palmer 1994) was
detected in the Z10.24 region. The SMA spectrum of the lower
transition at Z10.24 shows that a significant amount of gas is in
absorption while the emission is present but red-shifted with
respect to the absorption feature, a typical P-cygni profile
suggesting an outflow in this region. The spectrum of the higher
H$_{2}$CO transition can be fitted with two emission Gaussian
components at 50 and 84 km s$^{-1}$ with an absorbing Gaussian at
75 km s$^{-1}$, which is consistent with the H66$\alpha$
transition (de Pree, et al. 1996). Excluding the possibility of
the expanding shell model, de Pree et al. argued that an ionized
outflow is likely centered at Z10.24. Our H$_{2}$CO observations
appear to favor their argument of a bipolar outflow from the UCHII
region.

Towards the NE continuum source, both the higher and lower
H$_{2}$CO transitions show that the majority of the gas in
absorption is red-shifted with respect to the mean systematic
velocity 65 km s$^{-1}$.

Towards NW, a broad ($\Delta {\rm V}=18\pm1.3$ km s$^{-1}$)
absorption ($-1.1$ Jy beam$^{-1}$) from the lower transition line
is detected at 71 km s$^{-1}$. The spectrum of the higher
transition shows no significant lines in either emission or
absorption.

\subsection{Kinematics}

Figs. 5 and 6 show the images of intensity weighted velocity (or
moment 1) of the H$_{2}$CO emission gas from both transitions. The
moment 1 maps were constructed with a cutoff of 8 $\sigma$ from
each of the channel images in the velocity range of 8-151 km
s$^{-1}$.

In the Sgr B2(M) region, the kinematical structure observed from
the lower transition (see Fig. 5) consists of the highly
red-shifted components 5$^{\prime\prime}$ south-east of the
compact core and a northeast-southwest arch structure. The
component M4 appears to be a fast moving compact component
(V$_{\rm LSR}\sim100$ km s$^{-1}$) ejected from the core. The
morphology of the northeast-southwest arch in Sgr~B2(M) from the
velocity field in the higher transition gas (see Fig. 6) appears
to be consistent with that observed in the lower transitions. A
velocity gradient is present south-east of M4 (shown in Fig. 6),
which appears to indicate a decelerating outflow. In the case of
Sgr B2(M), the higher H$_{2}$CO transition appears to trace
outflow well, which is consistent with the interpretation of the
larger scale mass outflow based on the lower angular-resolution
observations of NH$_{3}$ and SO (Vogel, Genzel \& Palmer 1987).
The higher angular-resolution observations of NH$_{3}$ (Guame \&
Claussen 1990) showed that the red-shifted emission is located
south of the F3 HII region, and the blue-shifted absorption is
located north of the red-shifted emission, showing a velocity
gradient in north-south direction. Based on their higher
angular-resolution observations of OH maser and NH$_{3}$, Guame \&
Claussen (1990) suggested that the north-south velocity gradient
can be explained by a rotating disk or a torus of material with an
extent of 2$\rlap{.}^{\prime\prime}$5 surrounding the Sgr B2(M)-F
cluster. Although the angular resolution in our observations is
not adequate to verify the kinematical model proposed by Guame \&
Claussen (1990), our observed arch-liked morphology of the H$_2$CO
gas shown in blue-shifted emission from northeast to southwest in
the larger scale also indicates that the gas is undergoing a
complicated infall process interacting with the outflow while the
gas is spiraling onto the core rather than being in simple free
fall.

In the Sgr B2(N) region, the lower transition map (see Fig. 5)
shows the red-shifted velocity located north of K2 and
blue-shifted velocity located south of K2 with a north-south
velocity gradient across K2. A similar velocity gradient has been
observed in the HC$_{3}$N line (Lis et al. 1993). These authors
argued that the north-south velocity gradient in Sgr B2(N) traces
rotation. The kinematics of the lower H$_{2}$CO transition in
emission gas of Sgr~B2(N) is likely dominated by the gas rotating
around the core. In addition to the north-south velocity gradient
from the higher transition H$_{2}$CO( $3_{21}-2_{20}$) map as
observed in the lower transition gas, an east-west velocity
gradient is also observed in Sgr~B2(N) (see Fig. 6). The outflow
in Sgr~B2(N) was observed in east-west direction (Lis et al.
1993). The higher angular-resolution observations of NH$_{3}$
(Gaume \& Claussen 1990) appeared to show a southeast-northwest
velocity gradient across K2. The southeast-northwest velocity
gradient in the larger scale from our observations appears to be
consistent with the kinematical structure observed in NH$_{3}$.
Based on the higher angular-resolution observations of NH$_{3}$,
ruling out a simple outflow/rotation model, Gaume \& Claussen
(1990) suggested that several kinematic components in outflow,
infall and rotation might be involved in the Sgr B2(N) core. The
southeast-northwest velocity gradient observed from the H$_{2}$CO
emission gas in Sgr~B2(N) appears to be caused by a combination of
rotation, infall and outflow. The angular resolution of our
observations is not adequate to discern the details of these
motions.

\section {Outflow and Infall}

In order to better understand the on-going astrophysical processes
in the star formation cores, we modeled the observed kinematics
and the ratio of the line intensities. In comparison to Sgr~B2(N),
the kinematics observed from Sgr~B2(M) appears to be relatively
simple and characterized by infall and outflow. Fig. 7 shows the
single field map of  H$_2$CO ($3_{21}-2_{20}$) centered at the
F-cluster. The major axis of the red-shifted outflow
(PA=158$^{\circ}$) can be drawn by connecting the continuum core
and the tips of the outflows (M1 and M5).

Along the major axis of outflow (PA=158$^{\circ}$), we have made a
position-velocity diagram (PV) from the higher H$_{2}$CO
transition line cube (see Fig. 8). The solid contours show the
emission and dashed contours indicate the absorption. This diagram
shows two distinct decelerating outflow components clearly. The
red-shifted component shows that the emission near the core
(position at 0\arcsec~in the vertical axis) starts with a high
velocity of 106 km s$^{-1}$ and the velocity declines to a
terminal velocity at 58 km s$^{-1}$ as the gas goes to the outer
region (20\arcsec away from the core). If we shifted velocity to
the terminal velocity 58 km s$^{-1}$ or in the system rest frame,
the decelerating red-shifted outflow is mainly located in the
first quadrant with the absolute velocity decreasing from $\sim$48
to 0 km s$^{-1}$. The third quadrant shows the decelerating
blue-shifted outflow component with deceleration from $\sim$50 to
0 km s$^{-1}$.

Appendix A describes the numerical calculations that have been
carried out to model the kinematical characteristics of a
decelerating outflow combined with spherically infall. With a
simple assumption of mass conservation and power law distributions
in  both outflow velocity and molecular density, the observed
kinematics of the outflow components can be well fitted to the
decelerating outflow model with an initial velocity of V$_0$=83 km
s$^{-1}$ at r$_0$=1\arcsec~(0.04 pc) from the core center with the
outflow opening angle of $\Theta=$30$^\circ$ and inclination of
$\varphi=$45$^\circ$ for the red-shifted component (the thick red
curve in Fig. 8). The blue-shifted component can be fitted with
V$_0$=33 km s$^{-1}$ at r$_0$=1\arcsec ~(0.04 pc),
$\Theta=$30$^\circ$ and $\varphi=$45$^\circ$ (the thick blue
curve). The dashed curves range initial velocity V$_0$ at
r$_0$=1$^{\prime\prime}$ given the same opening angle and
inclination  angle. For the red-shifted outflow, V$_0$ is from 250
km s$^{-1}$ to 17  km  s$^{-1}$. The blue-shifted velocity ranges
from  83 km s$^{-1}$ to 4  km s$^{-1}$.

Our observations and analysis have shown the presence of a
decelerating outflow from the core in Sgr B2(M). Such a
decelerating outflow can be caused by the entrainment of gas
through the interaction between a fast wind flow from the central
stars and an ambient cloud core, where the incorporation of
additional mass into the outflow decelerates the wind (Lizano et
al. 1988). As the wind moves farther from the cores, the outflow
velocity terminates at the systematic velocity 58 km s$^{-1}$ of
the molecular cloud. We note that both the red-shifted and
blue-shifted outflows terminate at a common velocity, which is
close to the mean systematic velocity 61 km~s$^{-1}$ of Sgr B2(M)
(Sutton et al. 1991). In fact, the common terminal velocity
58$\pm$2 km s$^{-1}$ of blue-shifted and red-shifted outflow
appears to provide an unambiguous method, independent of the
chemical processes in molecular clouds, for accurately measuring
the systematic velocity of Sgr B2(M). The uncertainty in the
terminal velocity is mainly due to the velocity resolution in the
PV diagram and the uncertainty in the model fitting process.

Finally, the absorption located in the second quadrant is shown to
be red-shifted with respect to the systematic velocity 58$\pm$2 km
s$^{-1}$. Thus, with the accurately determined systematic velocity
we can be certain that this absorption feature arises from gas
flowing towards the continuum source.

\section{Excitation}

Based on the line ratio of the two transitions from the spectra,
most of the gas in the Sgr B2 region does not satisfy the LTE
condition. Under the assumption that the molecular cloud has
spherical geometry, the non-LTE calculation requiring the
collisional excitation rates for H$_2$CO (Green 1991) has been
carried out for Sgr B2 using the large velocity gradient (LVG)
approximation (e.g. Evans, Davis \& Plambeck 1979; Scoville \&
Solomon 1974; Lucy 1971). The radiative transfer calculations and
model fitting are discussed in details in Appendix B. The fitting
results for the line components for models with various sets of
parameters are given in Table 3.

We summarize the excitation properties for the various ${\rm
H_2CO}$ components in the following:

\noindent 1. Towards the absorption components of the cores (K1-3,
F1-4), the kinetic temperature depends on the beam filling factor
of the continuum emission, ranging from a few tens K for a large
continuum beam $f_{\rm C}$ (0.3) to a few hundreds K for a small
$f_{\rm C}$ (0.05). High angular resolution observations are
necessary to precisely determine the range of T$_{\rm k}$. The
typical values of ${\rm n_{H_2}}$ are 10$^{5-7}$ cm$^{-3}$ for the
core components. The derived column densities ${\rm
N_{para-H_2CO}/\Delta V}$ are in the range of 10$^{14-17}$
cm$^{-2}$ km$^{-1}$ s.

\noindent 2. For the discrete ${\rm H_2CO}$ components, no LVG
solutions could be fitted to the observations with either model A
or model B in which $f_{\rm C}$ is large and the brightness
temperature of the background radiation is too small to produce
significant absorption lines as observed. For excitation
temperature of a few tens K and higher, the intensity of
absorption (see equation(5)) suggests that the beam filling factor
must be small, {\it i.e.} $f_{\rm C}\sim$0.1 or smaller, for a
region with a continuum peak intensity of $\sim$2 Jy beam$^{-1}$
in our observations. The kinetic temperatures derived from the LVG
fitting are all below 100 K for both emission and absorption
components except for M3 emission component in which T$_{\rm
k}=150_{-50}^{+250}$ K. In general the components outside the
cores are cooler. The typical values of ${\rm n_{H_2}}$ are
10$^{4-6}$ cm$^{-3}$ for the discrete components. The derived
column density ${\rm N_{para-H_2CO}/\Delta V}$ is in the range of
10$^{13-16}$ cm$^{-2}$ km$^{-1}$ s.

\noindent 3. For the components in the red-shifted outflow of Sgr~
B2(M), the line ratio ${\displaystyle {\rm I}(3_{03}-2_{02})\over
 \displaystyle {\rm I}(3_{21}-2_{20})}$ is
significantly smaller than unity, suggesting that the population
of the ground levels in the two K ladders ($K_{-1}=0$ and 2) is
inverted. Hence, a weak maser process between the two different K
ladders must occur in this region. Based on our radiative transfer
code with the collisional excitation rates for kinetic temperature
T$_{\rm k}$$\le$ 300 K (Green 1991), we cannot rule out the
possibility that the H$_2$CO ($3_{21}-2_{20}$) line in Sgr B2(M)
outflow is excited in the high temperature C-shocks. On the other
hand, the strong FIR radiation field near the Sgr B2(M) core might
also play a role in the population inversion of the molecule.

\section{Discussion}

The assessment of molecular cloud mass from molecular lines can be
affected by the excitation, opacity, abundance variations and gas
dynamics of the molecular lines. The optically thin submillmeter
dust continuum emission has been proven to be a good tracer of
molecular cloud mass (Pierce-Price et al. 2000, Gordon 1995). If
we take an average grain radius of 0.1 $\mu$m and grain density of
3 g~cm$^{-3}$ and a gas to dust ratio of 100 (Hildebrand 1983,
Lis, Carlstrom \& Keene 1991), the dusty cloud mass and column
density are given by the formulae (Lis, Carlstrom \& Keene 1991)

 \begin{equation}
 M_{H_{2}}=1.3\times 10^{4}\frac{e^{h{\nu}/{\kappa}T}-1}{Q({\nu})}
 (\frac {S_{\nu}}{Jy})(\frac {D}{kpc})^{2}(\frac {{\nu}}{GHz})^{-3}~M_{\odot},
\end{equation}

\begin{equation}
N_{H_{2}}=8.1\times
10^{17}\frac{e^{h{\nu}/{\kappa}T}-1}{Q({\nu}){\Omega}}(\frac {
S_{\nu}}{Jy})(\frac {{\nu}}{GHz})^{-3}~ ({\rm cm}^{-2}),
\end{equation}

\noindent where $T$ is the mean dust temperature (K), $Q({\nu})$
is grain emissivity at frequency ${\nu}$, $S_{\nu}$ is the flux
density corrected for free-free emission, $\Omega$ is the solid
angle subtended by the source. Assuming $Q({\nu})$ at 1.3 mm is
2$\times$10$^{-5}$ and the dust temperature is 150 K for Sgr B2 (
Carlstrom \& Vogel 1989; Lis et al. 1993; Kuan, Mehringer \&
Snyder 1996), we derived the masses, H$_{2}$ column densities and
number densities.

Because the continuum at 1.3 mm contains free-free emission, we
estimate the physical parameters using the flux densities of the
continuum corrected for free-free emission. Assuming 3.6 cm
continuum emission with FWHM beam of
3$\rlap{.}^{\prime\prime}8\times 2\rlap{.}^{\prime\prime}0$
(Mehringer et al. 1993) of Sgr B2(N) (K1-3)  and Sgr B2(M) (F1-4)
come from optically thin free-free emission ($S_{\nu} \varpropto
{\nu}^{-0.1}$), we estimate the free-free contribution of 4.7 and
8.4 Jy ($\sim$ 9\% and  24\% of the total flux densities at 1.3
mm) at 1.3 mm towards the continuum cores K1-3 and F1-4 in our
observations. Our estimates are consistent with the determinations
of Lis et al. (1993) (6\% and 33\% of the total flux densities at
1.3 mm) and Martin-Pintado et al. (1990) ($\sim$ 9\% and 28\% of
the total flux densities at 1.3 mm ) for K1-3 and F1-4. The
continuum flux densities at 1.3 mm corrected for free-free
emission are 47.4 and 27.2 Jy for K1-3 and F1-4, respectively.

From the flux densities at 1.3 cm (Gaume et al. 1995), the
estimated free-free emission contributions at 1.3 mm are 0.02 Jy,
0.04 Jy and 1.26 Jy for K4, Z10.24 and MW, respectively. The
corresponding continuum flux densities corrected for free-free
emission are 3.5, 9.1 and 11.0 Jy, respectively.

The derived H$_{2}$ masses, column densities and number densities
are summarized in Table 4. The estimated H$_{2}$ masses of the Sgr
B2(N) core (K1-3) and the Sgr B2(M) core (F1-4) are larger than
those given by Lis et al. (1993), while the H$_{2}$ number
densities are less than their results. This result is caused by
the relatively larger size and higher flux densities of the
continuum in our observations.

The H$_{2}$CO ($3_{03}-2_{02}$) spectra show absorption against
both Sgr B2(N) and Sgr B2(M) compact cores and multiple absorbing
peaks. The absorptions are dominated by red-shifted gas,
suggesting that the lower transition H$_{2}$CO ($3_{03}-2_{02}$)
traces the cold gas in front of the continuum cores falling into
the two compact cores.

Previous observations showed multiple massive young stars in Sgr
B2(N) and Sgr B2(M) (e.g., Gaume et al. 1995, de Pree, Goss \&
Gaume 1998). There are multiple absorbing peaks with different
optical depths from our H$_{2}$CO spectra in the Sgr B2(N) and Sgr
B2(M) cores, which appears to indicate that the gas is falling
into the massive stars or massive star forming cores embedded at
different depths in the molecular clouds (Mehringer, Palmer \&
Goss 1995). However, the angular resolution of our observations is
inadequate for us to determine whether there are multiple regions
present or whether the overall gravitation potential dominates the
infalling gas. If infalling gas is in simple free-fall, the
infalling velocities can be estimated by

 \begin{equation}
 V_{in}=\sqrt {\frac {2MG}{r_{in}}}=0.09\sqrt {\frac
 {M/M_{\odot}}{r_{in}/pc}} ~ {\rm km~s^{-1}},
\end{equation}

\noindent where $r_{in}$ is the infall radius, $M$ is the sum of
the gas and star masses included in the $r_{in}$ and $G$ is the
gravitation constant.

The H$_{2}$ masses derived from the continuum are
1.4$\times$10$^{4}$ and 7.9$\times$10$^{3}$ M$_{\odot}$ for the
cores of Sgr B2(N) and Sgr B2(M), respectively. The VLA
observations of radio continuum at 1.3 cm (Gaume et al. 1995)
showed that there are three UCHII regions (K1, K2 and K3) in the
core of Sgr B2(N) and four UCHII regions (F1, F2, F3 and F4) in
the core of Sgr B2(M). By use of the relationships between stellar
spectral type and stellar mass (Vacca, Garmany \& Shull 1996), a
total stellar mass of 68 M$_{\odot}$ was inferred for the massive
stars in the core of Sgr B2(N). The higher resolution observations
(0$\rlap{.}^{\prime\prime}$05) at 7 mm (de Pree, Goss \& Gaume
1998) resolved out F1, F2, F3 and F4 into twenty-one UCHII
regions, and a stellar mass of 443 M$_{\odot}$ was inferred
corresponding to the massive stars in the core of Sgr B2(M). The
mass of the massive stars in the Sgr B2(M) core is six times
larger than that in the Sgr B2(N) core. Taking the major axis
sizes of 0.24 and 0.29 pc as the infall radii of the two cores
(K1-3 and F1-4), we inferred the infall velocities of 21 and 15
km~s$^{-1}$ for Sgr B2(N) and Sgr B2(M), respectively. Hence,
based on our SMA observations, we have shown that high-density
molecular gas is continuously feeding onto the active star
formation cores in both Sgr B2(M) and Sgr B2(N).

\section{Summary}

The continuum emission and H$_{2}$CO spectral lines at 1.3 mm were
observed with the SMA. We detected the continuum emission from the
compact cores of Sgr B2(N) and Sgr B2(M). Outside of the two
compact cores, a few clumps, K4, NE, NW, ME, MW and the filament
of Z10.24 were also observed. Except for the three newly observed
objects, NE, NW and ME, all of the other compact cores, clumps and
the filament have radio counterparts at centimeter wavelengths.

The spectra in the two H$_{2}$CO lines towards the continuum and
outside of the continuum showed the gas components in either
emission or absorption. From the integrated line intensity and
intensity weighted velocity maps in line emission, we identified
possible outflow motion in Sgr B2(N) and Sgr B2(M) cores. The
infalling gas was detected by the red-shifted absorbing gas
against the continuum in Sgr B2(N) and Sgr B2(M) compact cores and
the nearby dust clumps and filament Z10.24. The absorbing gas in
Sgr B2(N) and Sgr B2(M) cores is red-shifted with respect to the
systematic velocities, suggesting that the gas in front of the
continuum cores is flowing onto the continuum cores.

In comparison to Sgr B2(N), the kinematics in Sgr B2(M) are
relatively simple. We interpreted the observed outflow and infall
in Sgr B2(M) by a model incorporating a spherically symmetric
inflow with a decelerating outflow. A decelerating outflow from
Sgr B2(M) was evident in our observations. We showed that both the
red-shifted and blue-shifted outflow components share a common
terminal velocity. From the outflow terminal velocity, we
determined the systematic velocity of 58$\pm$2 km s$^{-1}$ for Sgr
B2(M). With the systematic velocity well determined in a manner
independent of chemistry in the molecular cloud, we are confident
that the majority of the absorbing gas at least in Sgr B2(M) is
red-shifted and flowing towards the active star forming core.

Using the observed two H$_{2}$CO lines, we have derived the
excitation conditions of Sgr B2 region in LVG approximation. For
the absorbing components in the Sgr B2(N) and Sgr B2(M) cores, the
kinetic temperature, the H$_2$ density and column density are in
the range of several tens to hundreds K, 10$^{5-7}$ cm$^{-3}$,
10$^{14-17}$ cm$^{-2}$ km$^{-1}$ s, respectively. The components
outside of the cores are relatively cooler, and the H$_2$ density
and column density are one order of magnitude less than those in
the Sgr B2(N) and Sgr B2(M) cores. In the Sgr B2(M) outflow
region, the intensity of the H$_{2}$CO (3$_{21}-2_{20}$) is larger
than that of the H$_{2}$CO (3$_{03}-2_{02}$), suggesting the
inversion of the distribution of the particle number between the
two ground levels of K$_{-1}$ =0 and 2. The H$_2$CO
($3_{21}-2_{20}$) might be excited by C-shocks in the outflow. The
radiative excitation by the strong FIR radiation field in Sgr
B2(M) may play an important role in the population inversion.

\acknowledgments
We thank the SMA staff for making the
observations possible. J.J. Wang acknowledges support from
National Natural Science Foundation of China (No. 10328306).

\appendix

\section{A Model for A Molecular Cloud With Inflow And Decelerating Outflow Components}

In this appendix, we discuss a model incorporating a spherically
symmetric inflow (Shu 1977) with a decelerating outflow (Cabrit \&
Bertout 1990, Raga et al. 1993). We calculated position-velocity
(PV) diagrams based on this infall-outflow model. These PV
diagrams can be compared with the observed PV diagrams from Sgr
B2(M) to better understand the physical processes in the Sgr B2(M)
molecular cloud.

\subsection{Infall}

For an ideal isothermal flow, under the assumption of spherically
symmetric collapse, the volume density $\rho(r)$ and velocity
profiles $V(r)$ as functions of radius $r$ within the collapsing
inner envelope can be described by the power-laws (Shu 1977),

\begin{equation}
 \rho(r)=\rho_{0}(\frac{r}{R})^{-\frac{3}{2}},~~~~V(r)=-(\frac
{2GM(r)}{r})^{\frac{1}{2}},
\end{equation}

\noindent where $R$ is outer boundary of the collapsing cloud in
spherical geometry, $\rho_{0}$ is the density at $R$, $M(r)$ is
the mass of the cloud interior to $r$, $G$ is the gravitational
constant. Because the actual core has a finite size, the infall is
terminated at an inner radius of  $r_{0}$. The negative sign in
the velocity indicates that the direction of the inflow is
opposite that of the unit vector of radius in the spherical
geometry. The geometry of a cloud with spherical inflow and
decelerating outflow is shown in the Cartesian coordinate system
($x$, $y$, $z$) in Fig. 9. The sky plane is the x-y plane.

The column density ($N$) at a sky position (x,y) can be calculated
by integrating the volume density $\rho(r)$. The column density in
the sky plane is

\begin{equation}
N(x,
y)=2\rho_{0}R\int_{0}^{\sqrt{1-a^2}}(s^{2}+a^{2})^{-\frac{3}{4}}
ds,
\end{equation}

\noindent  where $s=z/R$, $a =\sqrt{(x/R)^{2}+(y/R)^{2}}$. In
order to get an analytical form for the column density, we express
$(s^{2}+a^{2})^{-\frac{3}{4}}$ in a Taylor expansion. Neglecting
the contribution from higher order terms since the interesting
region corresponds to $a<<1$ (or $\sqrt{x^2+y^2}<<R$) in the
practical case, the final
solution of equation (A2) is

\begin{equation}
N(x, y) \cong 5\rho_{0} \frac{R^{3/2}} {(x^{2}+y^{2})^{1/4}}, ~~
~~~\sqrt{x^{2}+y^{2}}\ne0.
\end{equation}

\noindent At (x,y)=(0,0), the inflow terminates at the inner
boundary $r_{0}$. Assuming the emission from the far-side gas is
blocked by the central compact object, the effective column
density along the line-of-sight is

\begin{equation}
N(0, 0) = 2\rho_{0} R[ (\frac{r_{0}}{R})^{-\frac {1}{2}} - 1].
\end{equation}

From equation (A1), excluding the mass of the central compact
object, the mass $M(r)$ of the spherically symmetric cloud within
$r$ can be derived by

\begin{equation}
M(r)=\int_{r_0}^{r}4\pi r^{2}\rho (r) dr=\int_{r_0}^{r}4\pi r^{2}
\rho_{0} (\frac{r}{R})^{-\frac{3}{2}}dr=\frac{8\pi \rho_{0}R^3}{3
}[(\frac{r}{R})^{\frac{3}{2}}- (\frac{r_0}{R})^{\frac{3}{2}}].
\end{equation}

\noindent The mass interior to $R$ is $\displaystyle M_R \approx
\frac{8\pi \rho_{0}}{3 }R^{3}$ if $R>>r_0$. Therefore the equation
(A5) can be re-written as

\begin{equation}
M(r) \cong M_R (\frac{r}{R})^{\frac{3}{2}}.
\end{equation}

The radial velocity of infall gas is the projection of the
velocity vector {\bf V} on $z$ axis. The radial velocity averaged
along the line-of-sight is zero at a position away from the
central position (x, y)=(0,0) due to the cancellation of the
velocity in a spherically-symmetric inflow. At the position (x,
y)=(0,0), ignoring the contribution of the emission from the
region behind the central object, the velocity at (x, y)=(0,0) in
front of the central object can be expressed by

\begin{equation}
   V_{infall}=\frac{\int_{r_{0}}^{R}V(r)dr}{\int_{r_{0}}^{R}dr}=-\frac
   {\int_{r_{0}}^{R}(\displaystyle\frac
    {2GM(r)}{r})^{\frac{1}{2}}dr}{\int_{r_{0}}^{R}dr}=-\frac{4\sqrt{2}}{5}
\sqrt{\frac {GM_{R}}{R}} \frac{1-(\frac{r_{0}}{R})^{\frac{5}{4}}}
{1-\frac{r_{0}}{R}}.
\end{equation}

We note that the sign convention in the spherically symmetric
description of infalling gas in front of the central object is
opposite to the convention of the radial velocity with respect to
the central object in the local standard rest (LSR) frame. In
addition, in the derivation above, the systematic velocity is
assumed to be zero.

\subsection{Decelerating Outflow}

Molecular outflows are observed in star formation regions. Various
models have been also proposed to interpret the observed outflow
morphology (e.g., Raga et al. 1993; Cabrit \& Berout, 1990). In
this appendix, a PV diagram for a decelerating flow is calculated.
The outflow velocity field and density are assumed to have
power-law distributions along the major axis of the flow (cf.
Cabrit \& Bertout 1990)

\begin{equation}
V(r)=V_{0}(\frac{r_{min}}{r})^{\alpha},~~~~\rho(r)=\rho_{0}(\frac{r_{min}}{r})^{\delta},
\end{equation}

\noindent where $\rho_{0}$ and $V_{0}$ are the volume density and
velocity at the inner radius $r_{min}$, respectively;
$r=\sqrt{x^{2}+z^{2}}$, the distance from the center, because the
flow is in the x-z panel, i.e., $y=0$. The geometry of the
blue-shifted outflow is illustrated in Fig. 9 by the cone in the
$x-z$ plane with an opening angle of 2$\Theta$. The inclination
angle of $\varphi$ is the angle between the z-axis and the axis of
the blue-shifted outflow. The angle between the north and x-axis
is the negative position angle of the blue-shifted outflow if we
take the convention of position angle in AIPS. The radius of an
outflow cross-section is equal to $r\cdot {\rm tan}(\Theta)$.
Given conservation of mass flux across a cross-section and
$0<\Theta <90^{\circ}$, the relationship of the power-law indices
$\delta$ and $\alpha$ can be derived

\begin{equation}
\delta=2-\alpha.
\end{equation}

\noindent The size of a cut across the major-axis of the outflow
along the line-of-sight ($z$-axis) can be expressed approximately
by,
\begin{equation}
 L_{Z} \approx {\rm tan}(\Theta) x / {\rm sin}^{2}(\varphi),
\end{equation}

\noindent if we take $\alpha=1$ and $\delta=1$ (Cabrit \& Bertout
1990), then the column density along the line-of-sight can be
calculated from

\begin{equation}
N(x)=\int \rho dz=\int_{0}^{L_{Z}} \rho_{0}(\frac{r_{min}}{r}) dz,
\end{equation}

\noindent the result of the integration is

\begin{equation}
    N(x)=\rho_{0}r_{min}g(\Theta,\varphi), ~~~g(\Theta,\varphi)=ln
(\frac {{\rm tan}\Theta+({\rm tan}^{2}\Theta+{\rm
sin}^{4}\varphi)^{\frac{1}{2}}}{{\rm sin}^{2}\varphi}).
\end{equation}

\noindent The column density depends on the radius of the inner
cross section ($r_{min}$), volume density $\rho_{0}$ at $r_{min}$,
the opening angle ($\Theta$) and inclination angle ($\varphi$). In
practice, the $\varphi$ and $\Theta$ are in the ranges of
$0<\varphi <180^{\circ}$ and $0<\Theta <90^{\circ}$, respectively.
A blue-shifted outflow corresponds to $0<\varphi <90^{\circ}$ and
$90^{\circ}<\varphi <180^{\circ}$ to a red-shifted outflow. Fig.
10 plots  $g(\Theta,\varphi)$ as function of $\Theta$ and
$\varphi$. $g(\Theta,\varphi)$ is a weak function of $\Theta$ and
$\varphi$. Given $\Theta=30^{\circ}$ and $\varphi = 45^{\circ}$, $g$
is close to unity.

The mean radial velocity of the outflow can be expressed as
\begin{equation}
V(x)=\frac {\int v(r){\rm cos}\varphi dz}{\int dz}= \frac
{V_{0}r_{min}f(\Theta,\varphi)}{x}, ~~~~f(\Theta,\varphi)=\frac
{g(\Theta,\varphi){\rm sin}^{2}\varphi {\rm cos}\varphi }{{\rm
tan} \Theta}.
\end{equation}

\noindent The radial velocity of the outflow is inversely
proportional to $x$ along the major axis of the outflow and is a
weak function ($f(\Theta,\varphi)$) of $\Theta$ and $\varphi$. As
noted in the calculation of the infall velocity, the sign of the
velocity used in the equation (A13) is opposite to the convention
of the radial velocity with respect to the central source. The
velocity of the central source with respect to the LSR, or
systematic velocity, is not included here.

Fig. 11 shows the plot of $f(\Theta,\varphi)$ as a function of
$\Theta$ and $\varphi$. In the range of $90^{\circ}<\varphi
<180^{\circ}$, $f(\Theta,\varphi)$ has a negative value
corresponding to a red-shifted outflow, while $0^{\circ}<\varphi
<90^{\circ}$ is for a blue-shifted outflow. Given
$\Theta=30^{\circ}$ and $\varphi = 45^{\circ}$, $f$ is about 0.6.
In addition, for given $\varphi$, $r_{min}$, and $V_0$, the
observed radial velocity decreases as $\Theta$ increases.

Assuming $V_{0}= 85$ km~s$^{-1}$ at $r_{0}=1^{\prime\prime}$ (0.04
pc), $\Theta=30^{\circ}$ and $\varphi = 45^{\circ}$, we can
calculate a PV diagram of an outflow (the thick solid contour in
Fig. 12). Fig. 12 shows a PV diagram for the parameter
${V_{0}r_{min}f(\Theta,\varphi)}$ in a range between 0.5 to 5.1 km
s$^{-1}$ pc assuming that the blue and red-shifted outflows are
symmetric around the central source, whose velocity is 58 km
s$^{-1}$ with respect to the LSR. In this diagram, we also
includes an absorption component of a spherical inflow with a mean
infall velocity of 8 km s$^{-1}$ with respect to the central
source. The PV-diagram (Fig. 12) shows the typical configuration
of a spherically inflow and decelerating outflows in an active
star formation region.

\section{Radiative Transfer with LVG approximation}

In order to model the excitation conditions and the physical
properties of the molecular cloud components observed in Sgr B2,
we solve for the radiation transfer in a multi-level system with
the large velocity gradient (LVG) approximation. In this model, a
molecular cloud is assumed to be spherically uniform at kinetic
temperature (${\rm T_k}$), column density per unit velocity
interval (${\rm N_{para-H_2CO}/\Delta V}$) and H$_2$ density
(${\rm n_{H2}}$) with a volume filling factor $f_{\rm L, V}$ of
the line emitting gas. The escape probability,
$\beta=(1-e^{-\tau})/\tau$, is used to account for photon
trapping. The volume filling factor $f_{\rm L, V}$ is incorporated
into the opacity calculation via the equation

\begin{equation}
\tau_0={\rm A_{21}\over 8\pi \nu^3}
 {{\rm X_{para-H_2CO} n_{H_2}}f_{\rm L, V}
 \over {\rm dv/dz}} (x_1 {g_2\over g_1} -x_2),
\end{equation}

\noindent where $\tau_0$ is the optical depth at the line center,
A$_{21}$ is the Einstein coefficient for spontaneous emission,
${\rm X_{para-H_2CO}}$ is the abundance of para-H$_2$CO, ${\rm
dv/dz}$ is the velocity gradient, ${ x_{\rm i}}$ and ${g_{\rm i}}$
are the fractional population and statistical weight of level i,
respectively. The column density per unit velocity interval at the
line center in the LVG model is given by

\begin{equation}
{\rm N_{para-H_2CO}/\Delta V}={{\rm X_{para-H_2CO} n_{H_2}} f_{\rm
L, V}\over{\rm dv/dz}}.
\end{equation}

\noindent Given ${\rm n_{H_2}}$, ${\rm T_k}$ and ${\rm
N_{para-H_2CO}/\Delta V}$, the local LVG model gives the radiative
temperature or the intensity of the line requiring as input the
collisional excitation rates. We adopted the collision rates
calculated by Green (1991). Green used the interaction potential
between H$_2$CO and He. Since He has twice mass of H$_2$,
collisional excitation due to H$_2$ should be 2.2 times more
effective than collisional excitation by He (Green 1991; Mangum \&
Wootten 1993). The collisional excitation rates from Green were
multiplied by this factor in our calculations. The accuracy in the
Green's calculation of the total collisional excitation rates is
$\sim$20\%.

Incorporating the Green's collisional excitation rates and the
volume filling factor discussed above, we used the radiative
transfer code in {\it Miriad} to solve for the radiative
intensities from the para-H$_2$CO for the two transitions
$3_{03}-2_{02}$ and $3_{21}-2_{20}$. In the line emission case,
the line intensity ratio of the two transitions is a good
indicator of the kinetic temperature in molecular clouds ({\it
e.g.} Mangum \& Wootten 1993). This ratio becomes strongly
dependent upon the volume density (${\rm n_{H_2}}$) when the
kinetic temperature is greater than the upper state energy of the
highest excitation transition (${\rm E_u}$). Given a column
density (${\rm N_{para-H_2CO}/\Delta V}$), both the intensity of
the lower transition I($3_{03}-2_{02}$) and the intensity ratio
(I($3_{03}-2_{02}$)/I($3_{21}-2_{20}$)) as a function of ${\rm
T_k}$ and ${\rm n_{H_2}}$ can be calculated. We calculate five
models with different sets of volume filling factor of the
para-H$_2$CO gas ($f_{\rm L,V}$), beam filling factor of the line
emission or absorption ($f_{\rm L}$) and beam filling factor of
the continuum emission ($f_{\rm C}$). In model A, B, C and D, the
line filling factors $f_{\rm L,V}=0.25$ and $f_{\rm L}=0.4$ are
assumed while $f_{\rm L,V}=1$ and $f_{\rm L}=1$ are assumed in
model E. The continuum filling factors of $f_{\rm C} = 0.3, 0.2,
0.1, 0.05$ and 0.05 are assumed in model A, B, C, D and E
respectively. Fig.13 shows the LVG results calculated with model C
for the component K1-3 at 72 km s$^{-1}$. The absorption line
intensities of the two transitions $3_{03}-2_{02}$ and
$3_{21}-2_{20}$ are plotted as function of the kinetic temperature
and H$_2$ density. The line intensity ratio
I($3_{03}-2_{02}$)/I($3_{21}-2_{20}$) is also shown (bottom
panel). In the high density and cold region (the right-bottom
corner), both transitions becomes optically thick and the
excitation temperature is much smaller than the brightness
temperature of the background continuum radiation and the line
ratio becomes unity. When the gas becomes hot and less dense
(left-top corner), the line ratio becomes large. Fitting the
observed results of the line intensity and the intensity ratio to
the LVG model, we can determine ${\rm T_k}$ and ${\rm n_{H_2}}$.
We find that the absorption case is different from the case of
emission, and the absorption line intensity of the lower energy
transition, I($3_{03}-2_{02}$), is a good indicator of the kinetic
temperature (${\rm T_k}$) of a cloud component while the line
intensity ratio I($3_{03}-2_{02}$)/I($3_{21}-2_{20}$) places a
strong constraint on the H$_2$ volume density ( ${\rm n_{H_2}}$)
(see the top panel of Fig. 14).

Fig. 14 shows examples of the LVG fitting with model C ($f_{\rm
L,V}=0.25$,  $f_{\rm L}=0.4$, and $f_{\rm L,V}=0.1$) to the
observed results from three typical regions. Top panel in Fig. 14
is the result showing the LVG model curves fitting to the observed
I($3_{03}-2_{02}$) and I($3_{03}-2_{02}$)/I($3_{21}-2_{20}$) for
the velocity component at 72 km s$^{-1}$ in the region K1-3 of Sgr
B2(N). The dashed lines are absorption in the units of Jy
beam$^{-1}$. The thick dashed line and two thin dashed lines in
light blue are the observed values of I($3_{03}-2_{02}$) and 1
$\sigma$ uncertainty. The solid lines are the intensity ratio
I($3_{03}-2_{02}$)/I($3_{21}-2_{20}$). The thick line and two thin
lines in red correspond to the observed ratio and 1 $\sigma$
uncertainty. The hatched region marks the solution ranges of
2.9$\times10^6$--7.9$\times10^6$ cm$^{-3}$ in  ${\rm n_{H_2}}$ and
${\rm T_k}=350-360$ K for given ${\rm log(N_{para-H_2CO}/\Delta
V)=16.4}$ cm$^{-2}$ km$^{-1}$ s. We searched for LVG solutions in
a wide range of $11.0-19.0$ cm$^{-2}$ km$^{-1}$ s in log(${\rm
N_{para-H_2CO}/\Delta V}$).

The bottom-left panel in Fig. 14 shows the velocity component at
68 km s$^{-1}$ in the F1-4 region of Sgr B2(M). Given ${\rm
log(N_{para-H_2CO}/\Delta V)=17.0}$ cm$^{-2}$ km$^{-1}$ s, two
possible solutions were found for this region, one with higher
temperature (T$_{\rm k}=$ 150 K) and lower density ${\rm
n_{H_2}}=4.6\times10^{3}$ cm$^{-3}$ and other with lower
temperature (T$_{\rm k}=$ 110 K) and higher density ${\rm
n_{H_2}}=1.2\times10^{4}$ cm$^{-3}$.

The bottom-right panel in Fig. 14 shows a LVG model to the
observed results from the isolated emission region (M3) at 53 km
s$^{-1}$. Given ${\rm log(N_{para-H_2CO}/\Delta V)=13.7}$
cm$^{-2}$ km$^{-1}$ s, T$_{\rm k}=90-240$ K and ${\rm
n_{H_2}}=0.79-2.3\times10^{5}$ cm$^{-3}$ were found. The large
uncertainties in T$_{\rm k}$ and ${\rm n_{H_2}}$ are due to
relatively weaker line emission.

\clearpage

\clearpage

\begin{deluxetable}{lllllrrr}

\tabletypesize{\scriptsize} \tablenum{1} \tablewidth{0pt}

\tablecaption{The SMA Measurements of Continuum Emission}
\tablehead{ \colhead{Source}& \colhead{RA(J2000)} &
\colhead{$\Delta$RA}& \colhead{DEC(J2000)} &
\colhead{$\Delta$DEC}& \colhead{Angular size}& \colhead{Peak
Intensity}&
\colhead{Flux Density}\\
&~~h~~m~~s~~&~
(${\prime\prime}$)&~~$~~{\circ}$~~${\prime}$~~${\prime\prime}$~~&~~
(${\prime\prime}$) &~~~~~and (PA)
~~~~~ &~~~(Jy~beam$^{-1}$)&(Jy)~~~~~~ \\
}

\startdata

K1-3$^a$&17 47 19.89 ~&$\pm$0.1&--28 22
17.6~&$\pm$0.3&6$\rlap{.}^{\prime\prime}8\times3\rlap{.}^{\prime\prime}5(-25^{\circ}$)
&29.2$\pm$2&52.1$\pm$4  \\

K4 & 17 47 20.02~&$\pm$0.1 &--28 22
04.7~&$\pm$0.1&5$\rlap{.}^{\prime\prime}2\times4\rlap{.}^{\prime\prime}8(-49^{\circ}$)
&1.4$\pm$0.1&3.5$\pm$0.3\\

NE&17 47 20.56~~&$\pm$0.5 &--28 22 14.4~~&$\pm$0.3
&5$\rlap{.}^{\prime\prime}9\times1\rlap{.}^{\prime\prime}4(-71^{\circ}$)
&1.2$\pm$0.1&2.5$\pm$0.3\\

NW&17 47 19.27~~&$\pm$0.3 &--28 22
11.8~~&$\pm$0.3&5$\rlap{.}^{\prime\prime}1\times4\rlap{.}^{\prime\prime}1(0^{\circ}$)
&1.0$\pm$0.1&2.3$\pm$0.2\\

Z10.24& 17 47 20.04~~&$\pm$0.3&--28 22 41.2~~&$\pm$0.4
&13$\rlap{.}^{\prime\prime}3\times4\rlap{.}^{\prime\prime}4(66^{\circ}$)
&2.0$\pm$0.2&9.1$\pm$0.8\\

F1-4$^b$&17 47 20.17~~&$\pm$0.2&--28 23 04.9~~&$\pm$0.3
&7$\rlap{.}^{\prime\prime}8\times4\rlap{.}^{\prime\prime}4(-58^{\circ}$)
&20.2$\pm$1.3&35.6$\pm$3 \\

ME&17 47 21.65~~&$\pm$0.6&--28 22
57.0~~&$\pm$0.7&4$\rlap{.}^{\prime\prime}1\times2\rlap{.}^{\prime\prime}6(73^{\circ}$)&0.58$\pm$0.15&1.0$\pm$0.3\\

MW&17 47 19.56~~&$\pm$0.8&--28 23
04.3~~&$\pm$0.6&12$\rlap{.}^{\prime\prime}8\times8\rlap{.}^{\prime\prime}3(-13^{\circ}$)
&1.7$\pm$0.2&12.3$\pm$1.3\\

\enddata

\tablenotetext{\it a}{Sgr B2(N) compact core (K1-3) was fitted
with a Gaussian source
(6$\rlap{.}^{\prime\prime}8\times3\rlap{.}^{\prime\prime}5$,
PA=$-24.8^{\circ}$) centered on K3 (RA(J2000)=17$^{\rm h}$47$^{\rm
m}$19$\rlap{.}^{\rm s}$90,
DEC(J2000)=$-28^{\circ}$22$^{\prime}$16$\rlap{.}^{\prime\prime}$9)
and two point sources at K2 (RA(J2000)=17$^{\rm h}$47$^{\rm
m}$19$\rlap{.}^{\prime\prime}$88, DEC(J2000)=
$-28^{\circ}$22$^{\prime}$18$\rlap{.}^{\prime\prime}$6) and at
(RA(J2000)=17$^{\rm h}$47$^{\rm m}$19$\rlap{.}^{\prime\prime}$95,
DEC(J2000)=
$-28^{\circ}$22$^{\prime}$12$\rlap{.}^{\prime\prime}$3). The flux
densities of the Gaussian source and the two point sources are
29.4$\pm$3.7, 19.1$\pm$1.5 and 3.6$\pm$1.0 Jy, respectively. The
total flux density of the compact core (K1-3) is 52.1$\pm$4 Jy.}

\tablenotetext{\it b} {Sgr B2(M) compact core (F1-4) was fitted
with a Gaussian source
(7$\rlap{.}^{\prime\prime}8\times4\rlap{.}^{\prime\prime}4$,
PA=$-58.1^{\circ}$) located within F cluster at
(RA(J2000)=17$^{\rm h}$47$^{\rm m}$20$\rlap{.}^{\rm s}$17,
DEC(J2000)=$-28^{\circ}$23$^{\prime}$05$\rlap{.}^{\prime\prime}$1)
and a point source centered on F3 (RA(J2000)=17$^{\rm h}$47$^{\rm
m}$20$\rlap{.}^{\rm s}$14,
DEC(J2000)=$-28^{\circ}$23$^{\prime}$04$\rlap{.}^{\prime\prime}$6).
The flux densities of the Gaussian and the point sources are
21.3$\pm$2.6 and 14.3$\pm$1.0 Jy, respectively. The total flux
density of the compact core (F1-4) is 35.6$\pm$3 Jy.}

\end{deluxetable}

\begin{deluxetable} {lrrrrrrrrr}

\tabletypesize{\scriptsize}

\tablenum{2}

\tablewidth{0pt}

\tablecaption{The SMA Measurements of  H$_{2}$CO Lines}
\tablehead{&&H$_{2}$CO&(3$_{03}-2_{02}$)&&&H$_{2}$CO&(3$_{21}-2_{20}$)\\
\colhead{Source} & \colhead{I$_{\rm p1}$}& \colhead{V$_{\rm
LSR1}$}& \colhead{$\Delta$V$_{1}$}& \colhead{$\tau_{1}$}&
\colhead{I$_{\rm p2}$}& \colhead{V$_{\rm LSR2}$}& \colhead{
$\Delta$V$_{2}$}& \colhead{$\tau_{2}$}\\

&Jy beam$^{-1}$&km~s$^{-1}$&km~s$^{-1}$&&Jy
beam$^{-1}$&km~s$^{-1}$&km~s$^{-1}$& }

\startdata

K1-3

     &5.9$\pm$0.1&49$\pm$0.5&8$\pm$0.5&\dots

     &11.3$\pm$0.2&46$\pm$0.5&18$\pm$0.5&\dots\\

     &--8.3$\pm$0.3&56$\pm$0.5&3$\pm$0.5&0.3$\pm$0.1

     &4.7$\pm$0.1&60$\pm$0.5&5$\pm$0.5&\dots&\\

     &--19.0$\pm$1.6&63$\pm$0.5&9$\pm$0.5&1.1$\pm$0.2

     &--8.7$\pm$0.1&65$\pm$0.5&4$\pm$0.5&0.4$\pm$0.1\\

     &--23.5$\pm$0.7&72$\pm$0.5&12$\pm$1.2&1.6$\pm$0.1

     &--16.6$\pm$0.1&73$\pm$0.5&7$\pm$0.5&0.8$\pm$0.1\\

     &--20.0$\pm$1.1&84$\pm$0.5&11$\pm$0.5&1.1$\pm$0.1

     &--4.0$\pm$0.1&82$\pm$0.5&4$\pm$0.5&0.1$\pm$0.1\\

     &18.5$\pm$0.1&96$\pm$0.5&7$\pm$0.5&\dots

     &19.6$\pm$0.1&92$\pm$0.5&7$\pm$0.5&\dots\\

     &7.5$\pm$0.1&106$\pm$0.5&8$\pm$0.5&\dots

     &12.1$\pm$0.5&105$\pm$0.5&10$\pm$0.5&\dots\\

     &15.4$\pm$0.2&121$\pm$0.5&15$\pm$0.5&\dots

     &7.8$\pm$0.2&116$\pm$0.5&15$\pm$0.9&\dots\\


K4&

     $<0.54^a$&\dots&\dots&\dots

     &$<0.48^a$&\dots&\dots&\dots\\

NE&

     --1.2$\pm$0.1&72$\pm$0.5&20$\pm$1.1&2.5$\pm$1.3

     &--0.5$\pm$0.1&71$\pm$0.9&18$\pm$2.2&0.5$\pm$0.2\\

NW&

     --1.1$\pm$0.1&71$\pm$0.5&18$\pm$1.3&2.5$\pm$1.4

     &$<0.48^a$&\dots&\dots&\dots\\

Z10.24&

     --1.0$\pm$0.1&70$\pm$1.1&24$\pm$2.6 &0.7$\pm$0.1

     &--0.3$\pm$0.1&74$\pm$7.0&15$\pm$9.5&0.2$\pm$0.1\\

     &1.2$\pm$0.1&90$\pm$0.5&11$\pm$1.1&\dots

     &0.3$\pm$0.1&87$\pm$6.5&13$\pm$8.5&\dots\\

F1-4

     &0.9$\pm$0.7&45$\pm$2.0&18$\pm$7.8&\dots

     &1.6$\pm$0.1&39$\pm$0.5&7$\pm$0.5&\dots\\

     &1.6$\pm$0.2&54$\pm$0.5&2$\pm$0.5&\dots

     &1.5$\pm$0.1&53$\pm$0.5&3$\pm$0.5&\dots\\

     &--6.1$\pm$0.4&58$\pm$0.5&3$\pm$0.5&0.4$\pm$0.1

     &--3.8$\pm$0.3&58$\pm$0.5&2$\pm$0.5&0.2$\pm$0.1\\

     &--7.6$\pm$0.7&62$\pm$0.5&4$\pm$0.5&0.5$\pm$0.1

     &--2.2$\pm$0.4&62$\pm$0.8&4$\pm$1.4&0.1$\pm$0.1\\

     &--16$\pm$0.1&68$\pm$0.5&10$\pm$0.5&1.6$\pm$0.1

     &--5.1$\pm$0.9&67$\pm$1.5&5$\pm$3.4&0.3$\pm$0.1\\

     &--10$\pm$0.3&76$\pm$0.5&3$\pm$0.5&0.7$\pm$0.1

     &--2.0$\pm$0.4&74$\pm$0.6&2$\pm$1.0&0.1$\pm$0.1\\

     &1.6$\pm$0.1&96$\pm$0.5&17$\pm$0.9&\dots

     &1.7$\pm$0.1&91$\pm$0.5&13$\pm$0.6&\dots\\

     &1.1$\pm$0.1&112$\pm$0.5&3$\pm$0.5&\dots

     &1.0$\pm$0.1&107$\pm$0.5&5$\pm$0.6&\dots\\


ME&

     $<0.76^a$&\dots&\dots&\dots

     &$<0.67^a$&\dots&\dots&\dots\\

MW&

     6.6$\pm$0.1&53$\pm$0.5&6$\pm$0.5&\dots

     &4.3$\pm$0.1&53$\pm$0.5&6$\pm$0.5&\dots\\

     &--0.7$\pm$0.1&70$\pm$0.9&19$\pm$2.4&0.5$\pm$0.1

     &--0.1$\pm$0.1&70$\pm$3.8&17$\pm$9.8&0.1$\pm$0.1\\

M1&

      6.7$\pm$0.4&52$\pm$0.5&11$\pm$0.5&\dots

     &4.0$\pm$0.1&52$\pm$0.5&11$\pm$0.5&\dots\\

     &--1.4$\pm$0.1&64$\pm$2.3&21$\pm$3.4&0.9$\pm$0.1

     &--1.0$\pm$0.1&64$\pm$1.2&12$\pm$2.1&0.6$\pm$0.1\\

M2&

      2.6$\pm$0.1&55$\pm$0.5&6$\pm$0.5&\dots

     &0.5$\pm$0.1&53$\pm$0.5&4$\pm$1.1&\dots\\

M3&

     1.9$\pm$0.1&53$\pm$0.5&7$\pm$0.5&\dots

     &0.8$\pm$0.1&55$\pm$0.5&9$\pm$0.9&\dots \\

M4&

      --2.1$\pm$0.1&51$\pm$0.5&7$\pm$0.5&0.8$\pm$0.1

     &$<0.48^a$&\dots&\dots&\dots\\

     &--2.2$\pm$0.1&65$\pm$0.7&14$\pm$1.5&0.9$\pm$0.1

     &5.9$\pm$0.1&66$\pm$0.5&10$\pm$0.5&\dots \\

     &--0.8$\pm$0.3&70$\pm$0.5&3$\pm$1.1&0.2$\pm$0.1

     &4.2$\pm$0.2&70$\pm$0.5&3$\pm$0.5&\dots\\

     &--1.6$\pm$0.3&75$\pm$0.5&3$\pm$0.6&0.5$\pm$0.1

     &$<0.48^a$&\dots&\dots&\dots\\

     &1.8$\pm$0.1&100$\pm$0.5&8$\pm$0.5&\dots

     &$<0.48^a$&\dots&\dots&\dots\\

M5

     &0.2$\pm$0.1&66$\pm$3.1&16$\pm$7.4&\dots

     &1.2$\pm$0.1&64$\pm$0.5&14$\pm$0.8 &\dots\\

\enddata

\tablenotetext{\it}{ $^a$ 3$\sigma$ limit.}

\end{deluxetable}

\begin{deluxetable}{lcclcllc}

\tabletypesize{\scriptsize} \tablenum{3} \tablewidth{-5pt}
\tablecaption{LVG Model Results} \tablehead{ \colhead{Source}& &
\colhead{V$_{\rm LSR}$} & \colhead{I$_{\rm p1}$/I$_{\rm p2}$}&
\colhead{Model\tablenotemark{\it a}}& \colhead{T$_{\rm k}$} &
\colhead{n$_{\rm H_{2}}$} &
\colhead{$\rm log( { N_{para-H_{2}CO}\over \Delta V})$} \\
& &(km s$^{-1}$) & & &~~(K)& (10$^{5}$~cm$^{-3}$)&
 (cm$^{-2}$ km$^{-1}$ s)\\

} \startdata

K1-3   &(a)\tablenotemark{\it b}&63&2.2$\pm$0.2&A&
61$^{+12}_{-11}$  &
            12.6$^{+6.1}_{-3.3}$

       &14.9--15.2 \\

       &   &  &           &B& 135$^{+5}_{-5}$  & 17.8$^{+10.4}_{-0.5}$

       &15.3--15.6 \\

       &   &  &           &C& 200$^{+280}_{-70}$  & 10.0$^{+10.0}_{-7.5}$

       &15.3--15.7 \\

       &   &  &           &D& 300$^{+150}_{-150}$  &0.22$^{+0.03}_{-0.09}$

       &16.5--17.6 \\

       &   &  &           &E& 255$^{+95}_{-95}$  &0.07$^{+0.02}_{-0.01}$

       &17.1--17.3 \\

       &(a)&72&1.4$\pm$0.1&A& 61$^{+4}_{-3}$   & 71$^{+245}_{-26}$

       &15.1--15.2 \\

       &   &  &           &B& 105$^{+25}_{-35}$  & 8.3$^{+5.9}_{-3.3}$

       &15.8--16.2 \\

       &   &  &           &C& 356$^{+12}_{-10}$  & 35$^{+124}_{-27}$

       &16.3--16.8 \\

       &   &  &           &D& 80$^{+120}_{-30}$  &0.25$^{+0.07}_{-0.09}$

       &17.3--17.4 \\

       &   &  &           &E& 105$^{+75}_{-50}$  &0.09$^{+0.01}_{-0.03}$

       &17.3--17.4 \\

       &(a)&84&5$\pm$0.3&A& 25$^{+4}_{-5}$    & 79$^{+36}_{-23}$

       &13.7--14.1 \\

       &   &  &           &B& 34$^{+5}_{-7}$  & 316$^{+1268}_{-158}$

       &13.8--13.9 \\

       &   &  &           &C& 30$^{+14}_{-14}$  & 316$^{+2195}_{-158}$

       &13.5--13.6 \\

       &   &  &           &D& 200$^{+20}_{-40}$  &0.025$^{+0.007}_{-0.008}$

       &17.2--17.3 \\

       &   &  &           &E& 200$^{+100}_{-100}$  &0.03$^{+0.01}_{-0.00}$

       &17.0--17.11 \\

NE     &(a)&72&2.4$\pm$0.5&C&25$^{+1}_{-1}$    &
           25.0$^{+476}_{-11}$

       &14.6--14.8 \\

       &   &  &           &D& 55$^{+1}_{-10}$  &6.3$^{+93.7}_{-6}$

       &14.8--15.8 \\

       &   &  &           &E& 55$^{+2}_{-10}$  &5.0$^{+26.6}_{-2.5}$

       &14.9--15.5 \\

Z10.24 &(a)&70&3.3$\pm$1.1&C& 26$^{+1}_{-1}$     &
         10$^{+30}_{-4}$

       &14.6--14.8 \\

       &   &  &           &D& 50$^{+8}_{-20}$  &2.0$^{+23.1}_{-1.9}$

       &14.8--15.6 \\

       &   &  &           &E& 56$^{+3}_{-17}$  &2.5$^{+97.5}_{-2.3}$

       &14.7--15.1 \\

       &(e)&90&4.0$\pm$1.3&C& 50$^{+47}_{-11}$   & 1.45$^{+0.54}_{-0.60}$

       &13.3--14.1 \\

       &   &  &           &D& \dots & \dots & \dots\\

       &   &  &           &E& 60$^{+120}_{-15}$  &0.79$^{+1.72}_{-0.69}$

       &13.0--13.9 \\

F1-4   &(a)&58&1.6$\pm$0.2&A& 77$^{+3}_{-3}$   &
         17.8$^{+23.9}_{-6.6}$

       &15.4--15.7 \\

       &   &  &           &B& 132$^{+2}_{-3}$ & 10$^{+22}_{-6}$

       &15.7--16.0 \\

       &   &  &           &C& 288$^{+5}_{-8}$ & 8.91$^{+41.1}_{-5.76}$

       &16.3--17.1 \\

       &   &  &           &D& 575$^{+10}_{-15}$ & 7.08$^{+8.72}_{-3.92}$

       &17.1--17.4 \\

       &   &  &           &E& 260$^{+10}_{-20}$ & 0.05$^{+0.01}_{-0.01}$

       &17.05--17.15 \\

       &(a)&62&3.4$\pm$0.6&A& 40$^{+26}_{-26}$   & 7.94$^{+31.8}_{-4.78}$

       &14.6--14.9 \\

       &   &  &           &B& 40$^{+20}_{-20}$ &  12.6$^{+87.4}_{-8.6}$

       &14.4--14.7 \\

       &   &  &           &C& 180$^{+30}_{-60}$ & 0.063$^{+0.078}_{-0.043}$

       &16.88--17.04 \\

       &   &  &           &D& 425$^{+125}_{-25}$ & 0.28$^{+0.22}_{-0.12}$

       &17.0--17.08 \\

       &   &  &           &E& 280$^{+60}_{-120}$ & 0.04$^{+0.01}_{-0.01}$

       &16.86--17.02 \\

       &(a)&68&3.1$\pm$0.5&A& 30$^{+9}_{-10}$   & 17.8$^{+39.7}_{-7.8}$

       &14.60--14.9 \\

       &   &  &           &B& 41$^{+9}_{-11}$ & 31.6$^{+168}_{-19}$

       &14.5--14.7 \\

       &   &  &           &C& 110$^{+60}_{-25}$ & 0.10$^{+0.06}_{-0.08}$

       &17.1--17.2 \\

       &   &  &           &D& 465$^{+35}_{-195}$ & 0.14$^{+0.48}_{-0.07}$

       &17.0--17.18 \\

       &   &  &           &E& 240$^{+80}_{-100}$ & 0.06$^{+0.02}_{-0.01}$

       &17.0--17.2 \\

       &(a)&76&5.0$\pm$1.0&A& 26$^{+13}_{-10}$   & 36.3$^{+122}_{-16.3}$

       &14.1--14.2 \\

       &   &  &           &B& 25$^{+21}_{-12}$   & 63.1$^{+137}_{-23.3}$

       &13.8--13.92 \\

       &   &  &           &C& 24$^{+1}_{-1}$     & 150$^{+166}_{-79}$

       &13.25--13.76 \\

       &   &  &           &D& 500$^{+30}_{-50}$     & 0.018$^{+0.032}_{-0005}$

       &16.8--17.1 \\

       &   &  &           &E& 150$^{+30}_{-70}$     & 0.02$^{+0.005}_{-0.005}$

       &16.8--16.89 \\

MW     &(e)&53&1.6$\pm$0.1&C& 34$^{+1}_{-1}$   &
            4.50$^{+1.81}_{-1.42}$

       &15.0--15.16 \\

       &   &  &           &D& \dots & \dots & \dots\\

       &   &  &           &E& 16$^{+1}_{-1}$     & 7.94$^{+7.86}_{-7.38}$

       &15.4--15.8 \\

       &(a)&70&3.5$\pm$1.7&C& 28$^{+1}_{-4}$   & 3.20$^{+12.6}_{-1.90}$

       &14.83--15.77 \\

       &   &  &           &D& 52$^{+8}_{-22}$     & 2.0$^{+77.4}_{-1.6}$

       &14.4--14.9 \\

       &   &  &           &E& 53$^{+6}_{-31}$     & 1.68$^{+48.4}_{-1.52}$

       &14.9--15.0 \\

M1     &(e)&52&1.7$\pm$0.1&C& 35$^{+5}_{-4}$   &
         3.47$^{+1.54}_{-1.07}$

       &15.00--15.16 \\

       &   &  &           &D& \dots & \dots & \dots\\

       &   &  &           &E& 15$^{+1}_{-1}$     & 10.0$^{+15.1}_{-9.44}$

       &15.3--16.8 \\

       &(a)&64&1.6$\pm$0.2&C& 25$^{+1}_{-1}$   & 3.47$^{+1.54}_{-1.07}$

       &15.05--15.1 \\

       &   &  &           &D& 55$^{+1}_{-4}$     & 20$^{+20}_{-9}$

       &15.2--16.1 \\

       &   &  &           &E& 90$^{+1}_{-1}$     & 12.6$^{+37.5}_{-11.3}$

       &15.2--16.0 \\

M2     &(e)&55&5.2$\pm$1.0&C& 40$^{+10}_{-7}$    &
        10.0$^{+5.8}_{-2.4}$

       &13.45--13.8 \\

       &   &  &           &D& \dots & \dots & \dots\\

       &   &  &           &E& 45$^{+35}_{-15}$     & 2.82$^{+7.18}_{-2.19}$

       &13.00--13.85 \\

M3     &(e)&53&2.4$\pm$0.3&C&150$^{+249}_{-50}$  &
        1.86$^{+0.77}_{-0.66}$

       &13.62--13.8 \\

       &   &  &           &D& \dots & \dots & \dots\\

       &   &  &           &E& 150$^{+250}_{-50}$     & 0.45$^{+0.81}_{-0.31}$

       &13.30--13.85 \\

\enddata

\tablenotetext{\it a}{The filling factors are $f_{\rm C}$=0.3,
$f_{\rm L}$=0.4, $f_{\rm L, V}$=0.25 for Model A, $f_{\rm C}$=0.2,
$f_{\rm L}$=0.4, $f_{\rm L, V}$=0.25 for Model B, $f_{\rm C}$=0.1,
$f_{\rm L}$=0.4, $f_{\rm L, V}$=0.25 for Model C , $f_{\rm
C}$=0.05, $f_{\rm L}$=0.4, $f_{\rm L, V}$=0.25
 for Model D and $f_{\rm C}$=0.05, $f_{\rm L}$=1, $f_{\rm L, V}$=1 for Model E.
For the absorption components, we used the following formula in
the radiative transfer calculation:
   $$\Delta {\rm T}_{\rm L}^{\rm obs} = (f_{\rm L} {\rm T}_{\rm ex} -{f_{\rm L}\over f_{\rm C}} {\rm T}_{\rm C}^{\rm obs})
(1- e^{-\tau_{\rm L}}).$$ For the emission components, we assumed
$f_0=0$ and used the following formula:
 $$\Delta {\rm T}_{\rm L}^{\rm obs} = (f_{\rm L} {\rm T}_{\rm ex} - {\rm T}_{\rm cmb}) (1- e^{-\tau_{\rm L}}),$$
 where ${\rm T}_{\rm cmb}$ is the temperature of
the cosmic microwave background radiation. }

\tablenotetext{\it b}{ The (a) and (e) indicate absorption and
emission, respectively.}
\end{deluxetable}

\begin{deluxetable}{lccc}

 \tabletypesize{\scriptsize} \tablenum{4} \tablewidth{-5pt}

\tablecaption{The Properties of Sgr B2 Derived from Dust Emission}

\tablehead{ \colhead{Source}& \colhead{M$_{\rm H_2}$} &

\colhead{N$_{\rm H_{2}}$} &

\colhead{n$_{\rm H_{2}}$} \\

&(10$^{3}$~M$_{\odot}$)& (10$^{24}$~cm$^{-2}$)&

 (10$^{6}$~cm$^{-3}$)

 \\

} \startdata

K1-3$^a$ &13.7$\pm$1&21.2$\pm$2&18.7$\pm$2 \\

K4 &1.0$\pm$0.1&1.5$\pm$0.1&1.3$\pm$0.1 \\

NE&0.73$\pm$0.06&3.2$\pm$0.3 &4.8$\pm$0.4\\

NW&0.66$\pm$0.06&1.2$\pm$0.1&1.1$\pm$0.1\\

Z10.24 &2.6 $\pm$0.2&1.7$\pm$0.1&0.9$\pm$0.1 \\

F1-4$^b$ &7.8$\pm$0.8&9.0$\pm$1.3&6.6$\pm$0.7\\

ME&0.28$\pm$0.03&1.0$\pm$0.1&2.4$\pm$0.3\\

MW&3.2$\pm$0.4&1.1$\pm$0.1&0.46$\pm$0.06\\

\enddata

\tablenotetext{\it a} {The core of Sgr B2(N).}

\tablenotetext{\it b} {The core of Sgr B2(M).}

\end{deluxetable}





\clearpage

\begin{figure}
\hspace{-10mm} \epsscale{1.0} \plotone{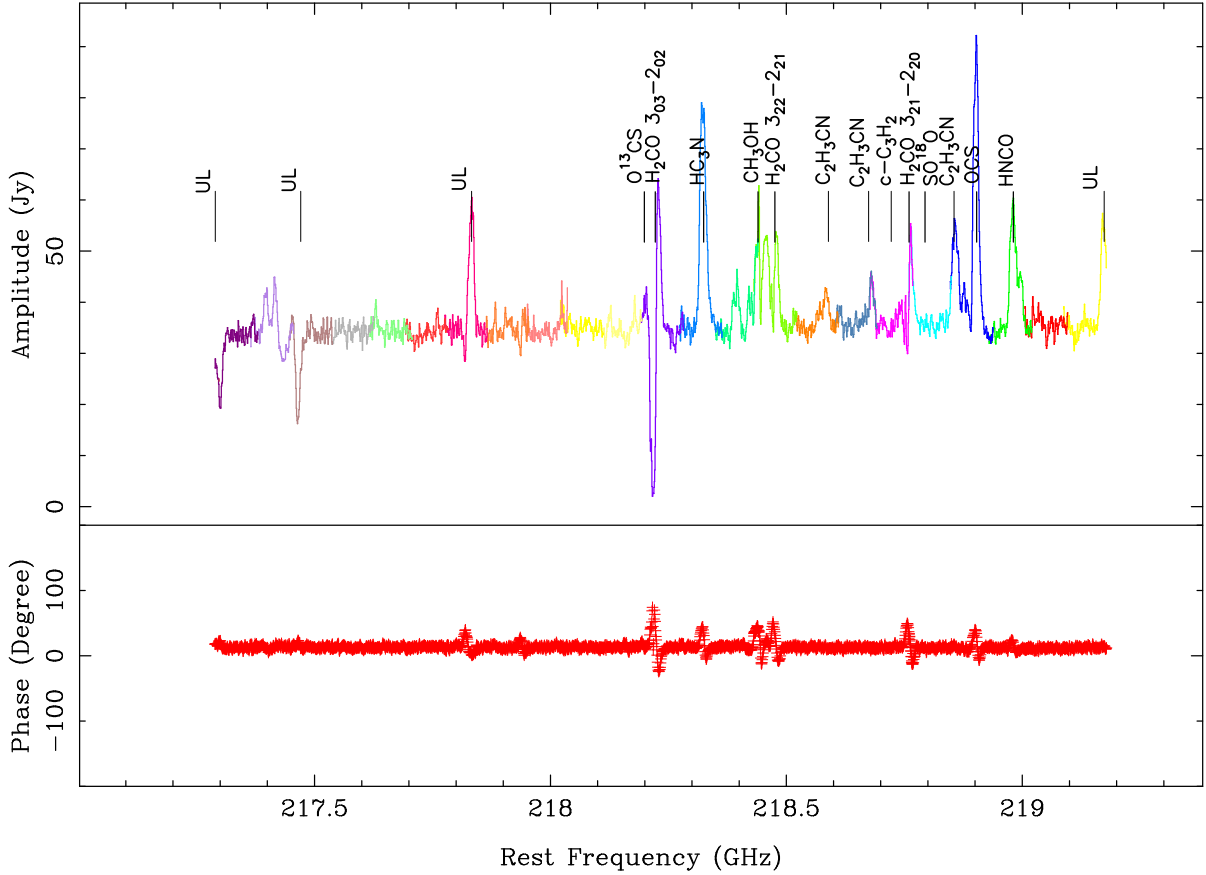} \caption{The
spectrum from (u, v) domain on the baseline 3-4 for Sgr B2(M). The
horizontal axis is the frequency in the rest frame and the
vertical is the amplitude and phase. The 24 spectral windows of
the SMA's spectrometer are coded in different color. UL indicates
the unidentified lines. We note that the features at 218.222 and
218.760 GHz are apparently dominated by H$_{2}$CO (3$_{03}-2_{02}$
and 3$_{21}-2_{20}$) and the possible contaminations from
O$^{13}$CS, c-C$_3$H$_2$ and SO$^{18}$O at the transitions near
the H$_{2}$CO frequencies 218.222 and 218.760 GHz appear to be
insignificant in Sgr B2(M). These molecular lines also appear to
be insignificant in Sgr B2(N). }

\end{figure}

\begin{figure}

\hspace{-15mm} \epsscale{0.8} \plotone{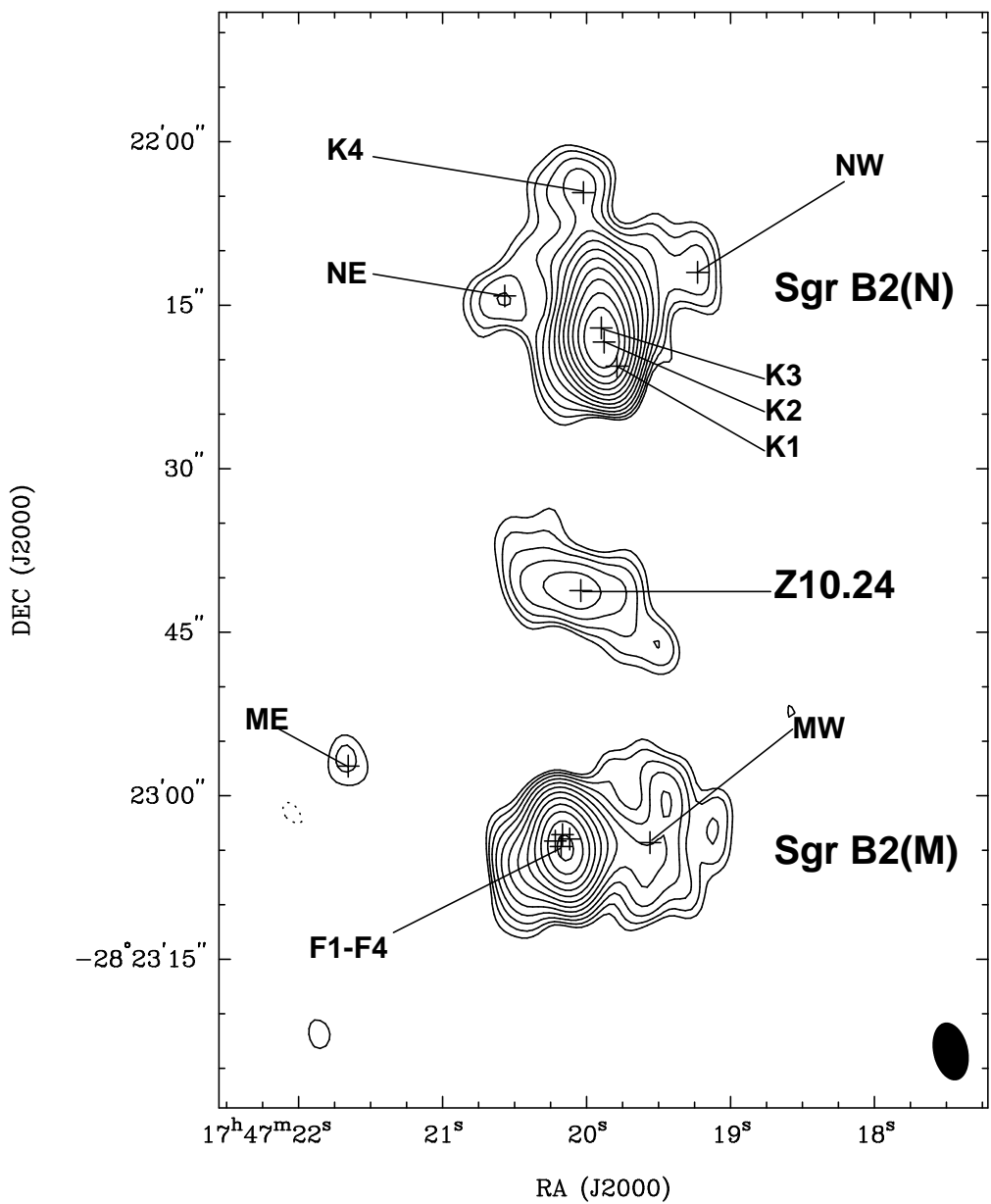} \caption{The mosaic
image of the continuum emission towards Sgr B2 at 1.3 mm obtained
by combining the data from both sidebands (218 and 228 GHz). The
synthesized beam is 5$\rlap{.}^{\prime\prime}4\times
3\rlap{.}^{\prime\prime}2$, PA=12.5$^{\circ}$ (lower-right
corner). The contours are -4, 4, 5.6, 8, 11, 16, 23, 32, 45, 64,
91, 128, and 181 $\sigma$. The rms (1 $\sigma$) noise level is 0.1
Jy~beam$^{-1}$. The plus symbols indicate the positions of the
continuum sources at 1.3 mm. The UCHII region K1-3 and F1-4
positions were taken from Gaume et al. 1995. }

\end{figure}

\begin{figure}

\hspace{-10mm} \epsscale{1.1} \plotone{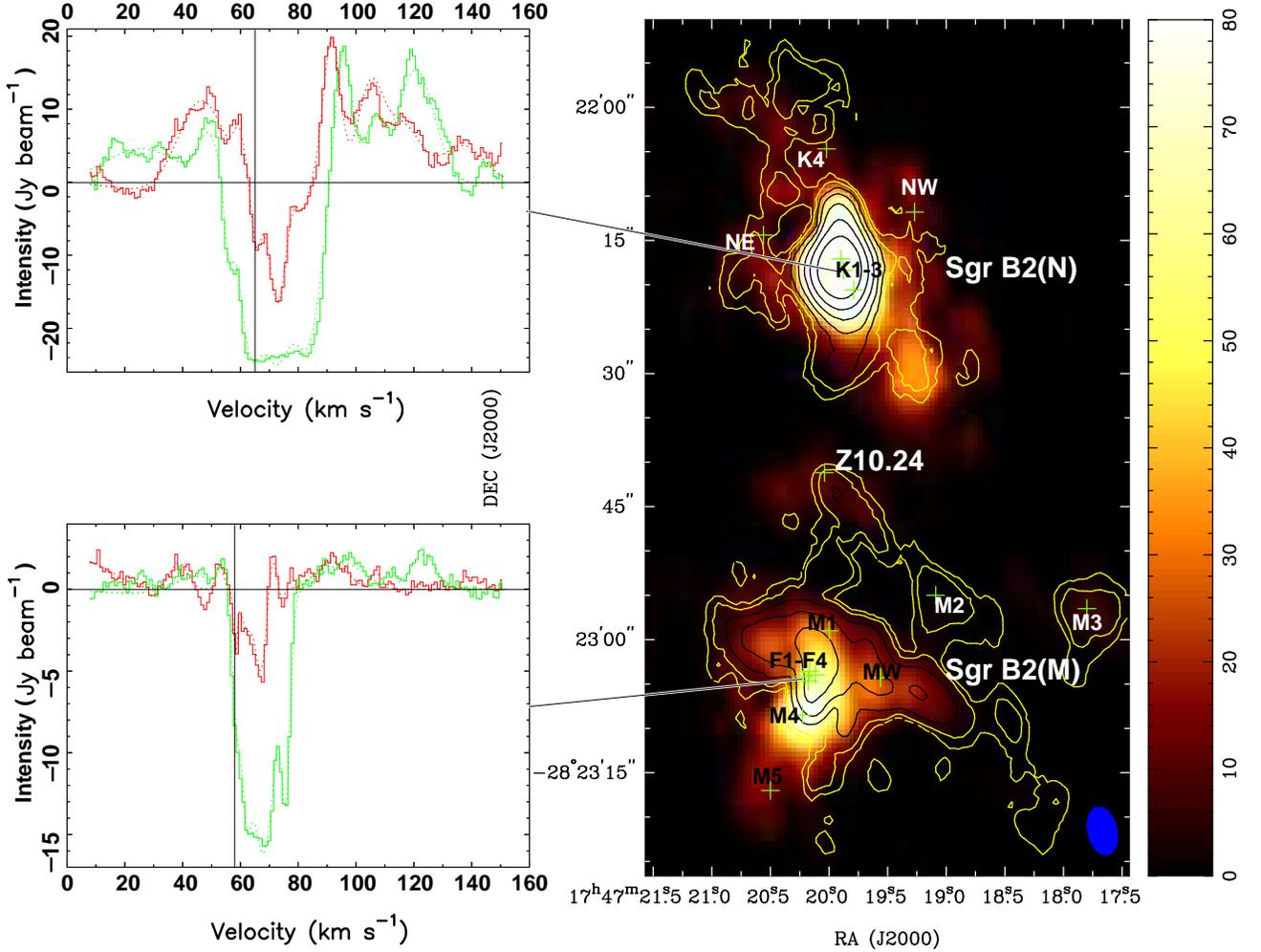} \vspace{-8mm}

\caption{The mosaic images of integrated intensities with FWHM
beam of 5$\rlap{.}^{\prime\prime}4\times
3\rlap{.}^{\prime\prime}2$, PA=12.5$^{\circ}$. The pseudo-color is
for integrated H$_{2}$CO ($3_{21}-2_{20}$) emission intensity with
4 $\sigma$ cutoff in each channel (1 $\sigma$  is 0.16
Jy~beam$^{-1}$). The color scale on the right shows the range from
0 to 80 Jy~beam$^{-1}~$km~s$^{-1}$. The contours (4, 8, 16, 32,
64, 128, 256 Jy~beam$^{-1}~$km~s$^{-1}$) are the integrated
H$_{2}$CO ($3_{03}-2_{02}$) emission intensity with 4 $\sigma$
cutoff in each channel (1 $\sigma$ is 0.18 Jy~beam$^{-1}$).
Top-left panel shows the observed spectra (solid curves) towards
Sgr B2(N), green for H$_{2}$CO ($3_{03}-2_{02}$) and red for
H$_{2}$CO ($3_{21}-2_{20}$). The dashed curves are the Gaussian
fitting to the multiple components. Bottom-left panel shows the
spectra towards Sgr B2(M). The vertical lines mark the systematic
velocities, 58 km s$^{-1}$ for Sgr B2(M) and 65 km s$^{-1}$ for
Sgr B2(N).}

\end{figure}

\begin{figure}

\epsscale{1.} \plotone{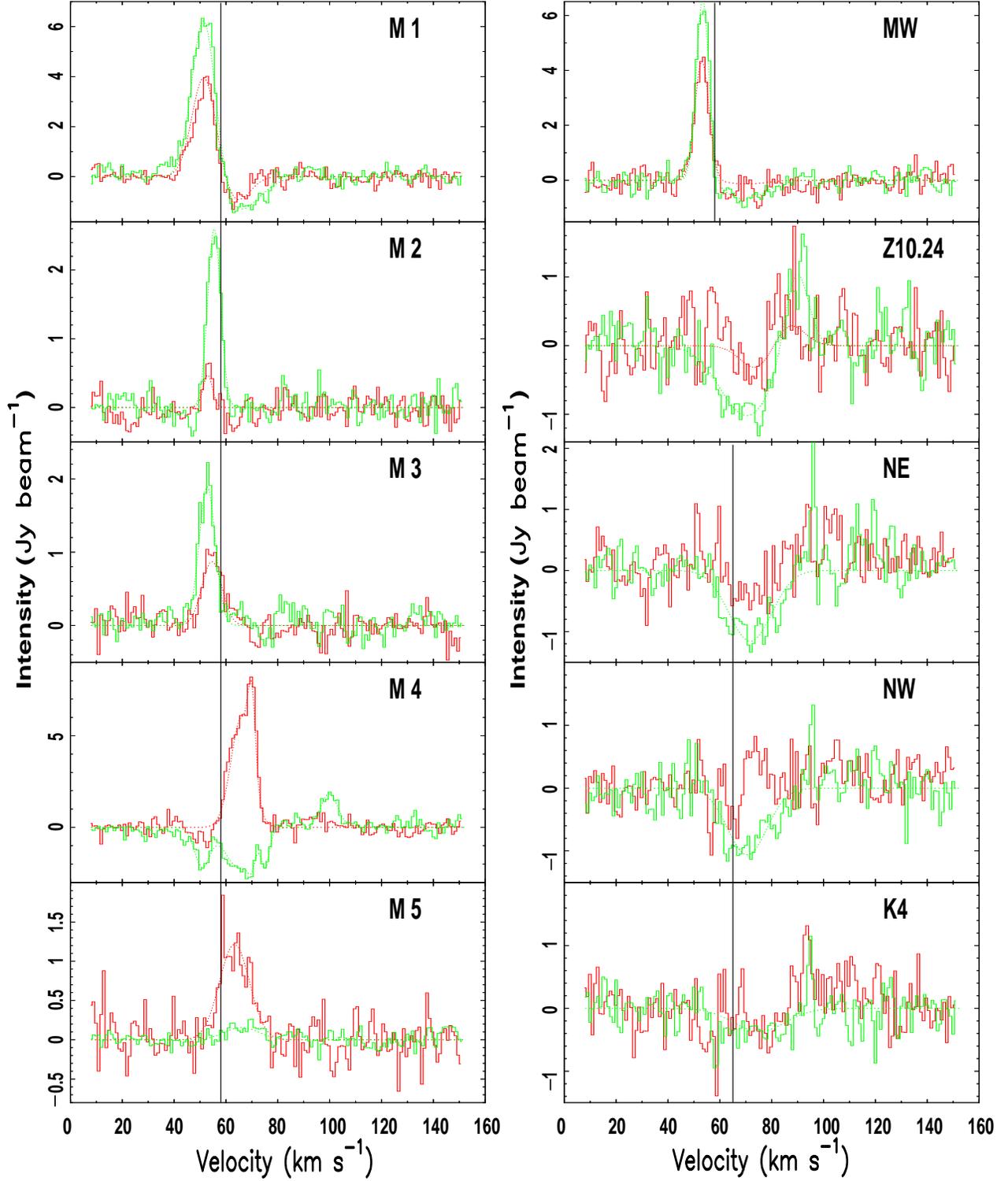} \vspace{-8mm} \caption{Spectra from
various components, green for H$_{2}$CO ($3_{03}-2_{02}$) and red
for H$_{2}$CO ($3_{21}-2_{20}$). Solid curves are the observed
spectra and dashed are the multiple Gaussian fitting. The vertical
lines mark the systematic velocities, 58 km s$^{-1}$ for M1, M2,
M3, M4, M5 and MW in Sgr B2(M) and 65  km s$^{-1}$ for NE, NW and
K4 in Sgr B2(N).} \label{fig:graphics_01}

\end{figure}

\begin{figure}

\epsscale{.80} \plotone{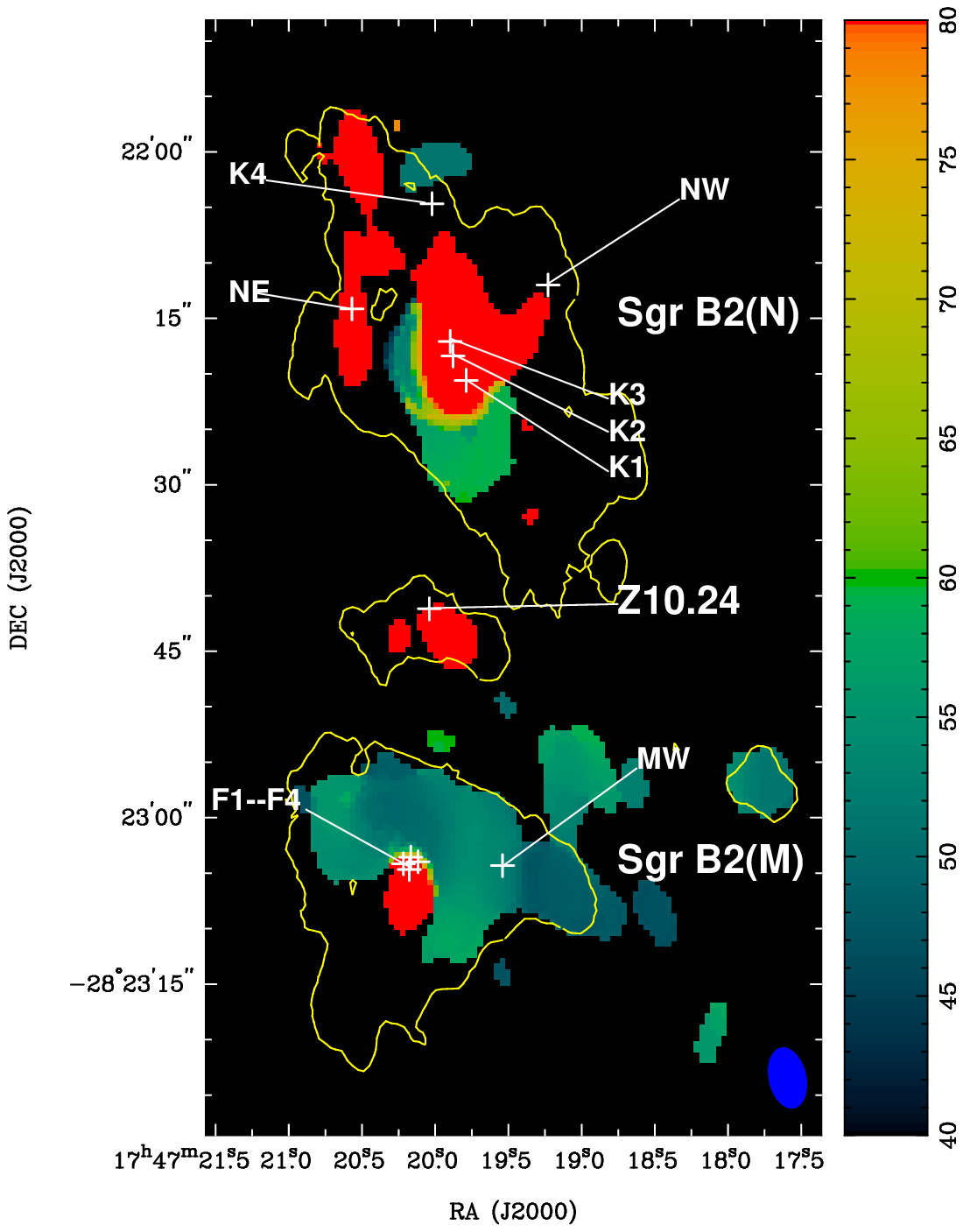} \caption{The mosaic image of the
intensity weighted velocity map in pseudo-color constructed from
the data after imposing a cutoff of 8 $\sigma$, showing velocity
field traced by H$_{2}$CO ($3_{03}-2_{02}$) emission in
pseudo-color. The scale on the right shows the velocity range from
40 to 80 km~s$^{-1}$. The contours outline the integrated
intensity of H$_{2}$CO ($3_{21}-2_{20}$) emission at the level of
4 Jy~beam$^{-1}$ km~s$^{-1}$. The FWHM beam is
5$\rlap{.}^{\prime\prime}4\times 3\rlap{.}^{\prime\prime}2$
(PA=12.5$^\circ$). }

\end{figure}

\begin{figure}

\centering \plotone{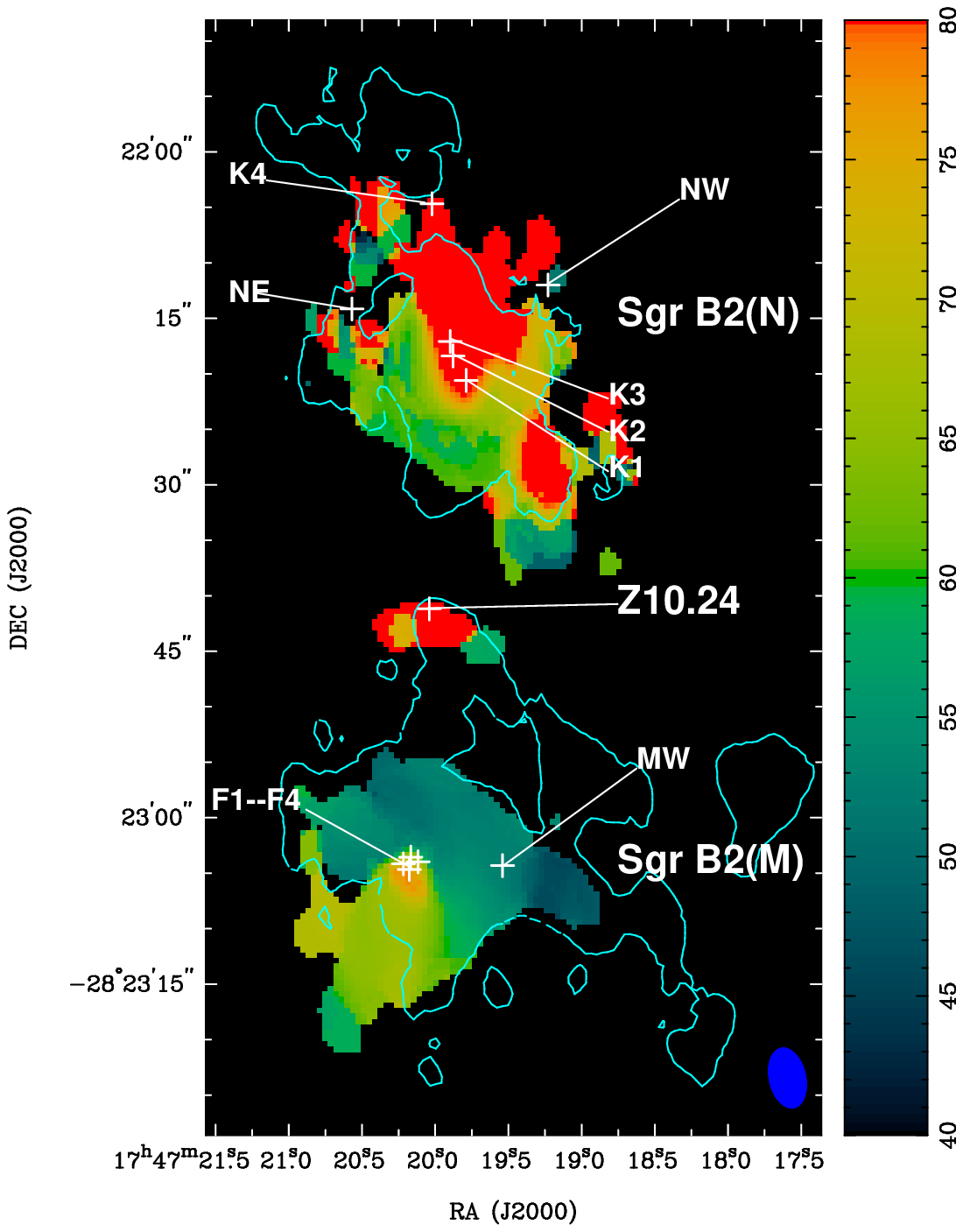} \caption{The mosaic image of the
intensity weighted velocity map in pseudo-color constructed from
the data after imposing a cutoff of 8 $\sigma$, showing velocity
field traced by H$_{2}$CO ($3_{21}-2_{20}$) emission in
pseudo-color. The  scale shows the velocity range from 40 to 80
km~s$^{-1}$. The contours outline the integrated intensity of
H$_{2}$CO ($3_{03}-2_{02}$) emission at the level of 4
Jy~beam$^{-1}$ km~s$^{-1}$. The FWHM beam is
5$\rlap{.}^{\prime\prime}4\times 3\rlap{.}^{\prime\prime}2$
(PA=12.5$^\circ$). } \label{fig:graphics_02}

\end{figure}

\begin{figure}
\epsscale{0.9}\hspace{-18mm} \plotone{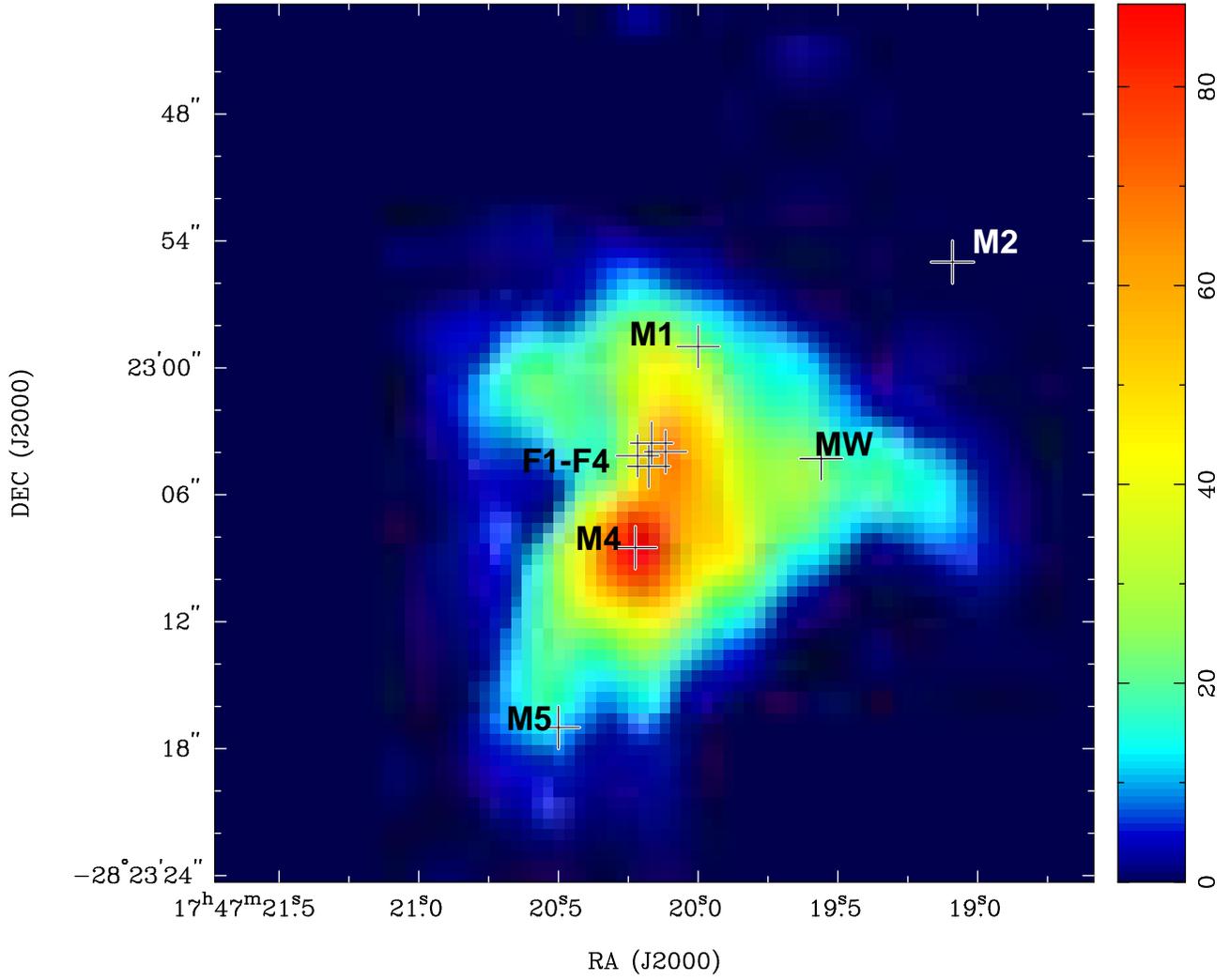}  \caption{The
integrated line intensity of H$_{2}$CO ($3_{21}-2_{20}$) in Sgr
B2(M), showing the emission from the decelerating outflow along
the direction in PA=158$^\circ$. The color scale gives the
intensity from 0 to 90 Jy beam$^{-1}$ km s$^{-1}$. The FWHM beam
is 5$\rlap{.}^{\prime\prime}4\times 3\rlap{.}^{\prime\prime}2$
(PA=12.5$^\circ$).}
\end{figure}

\begin{figure}

\centering \epsscale{.80} \plotone{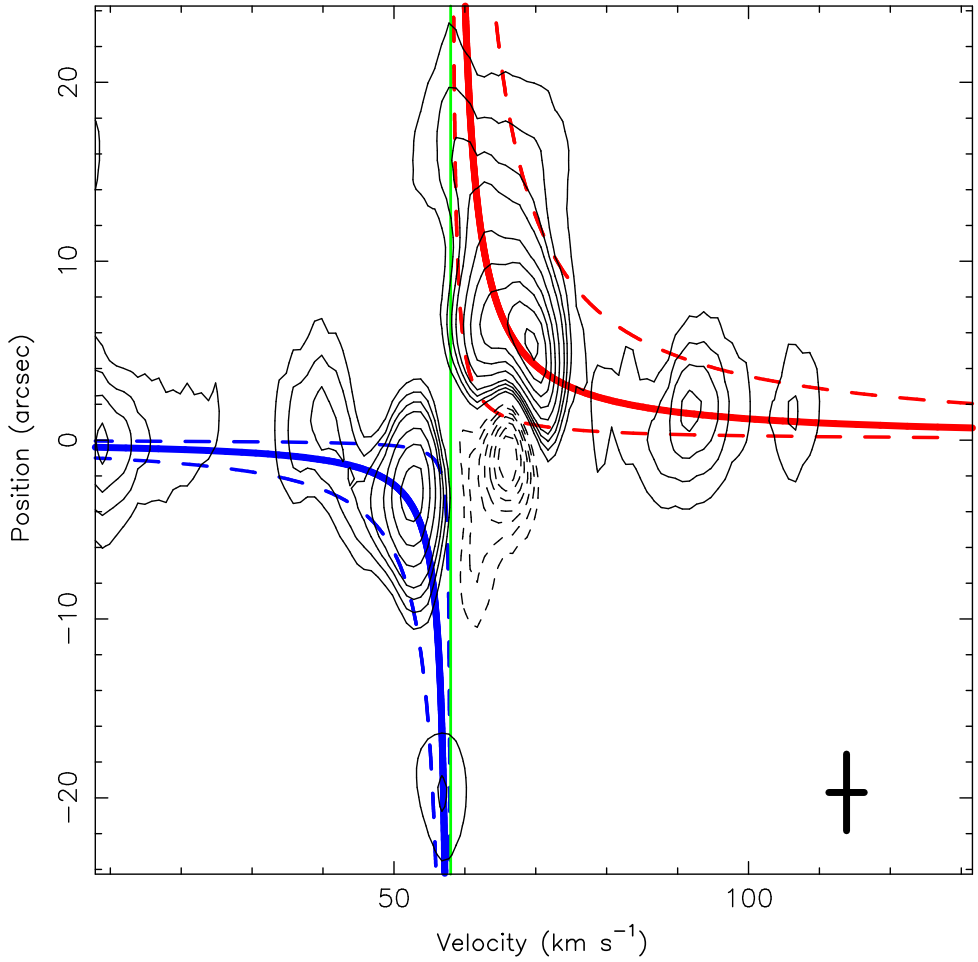} \caption{The
velocity-position diagram cutting along the major axis of the
outflows (see Fig. 7), which was constructed from the H$_{2}$CO
(3$_{21}-2_{20}$) spectral line cube. The solid contours show the
emission from the decelerating outflows. The dashed contours are
the absorption. The effective resolution of this map is labelled
with the plus sign at bottom-right corner, i.e. 5 km s$^{-1}$ in
velocity and 5 arcsec in position. The vertical green line marks
the systematic velocity. The solid curves are the best fitting to
the decelerating outflow, with the decelerating outflow parameter
$V_0 r_{min} f(\Theta,\varphi)=2$ km s$^{-1}$ pc for the
red-shifted component (red) and $V_0 r_{min}
f(\Theta,\varphi)=0.8$ km s$^{-1}$ pc for the blue-shifted. The
dashed color lines show the range of  $V_0 r_{min}
f(\Theta,\varphi)$ between 0.4 to 6 km s$^{-1}$ pc for the
red-shifted component and 0.1 to 2 km s$^{-1}$ pc for the
blue-shifted component. }

\end{figure}

\begin{figure}

\centering \centering \epsscale{.70} \plotone{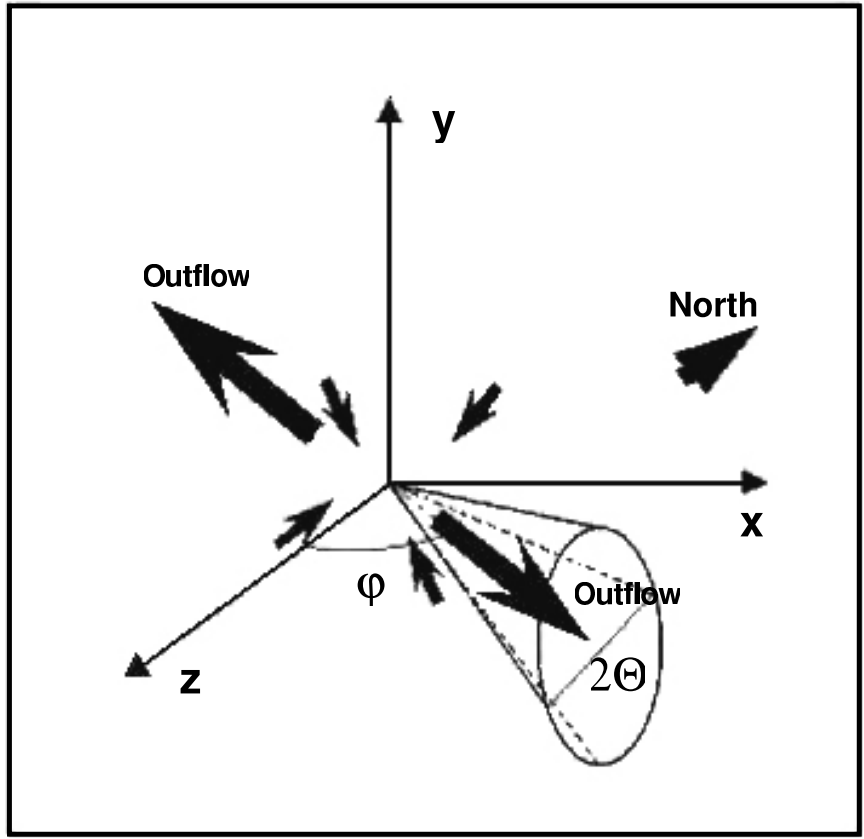} \caption{The
geometry of the layout for a spherical infall (indicated by the
small arrows towards the origin of the coordinates) and a
decelerating outflow. Here the sky plane is the $x-y$ plane. The
outflow lies in the $x-z$ plane with an opening angle of $2\Theta$
and an inclination angle of $\varphi$. The z-axis points towards
the observer. }

\end{figure}

\begin{figure}

\centering \plotone{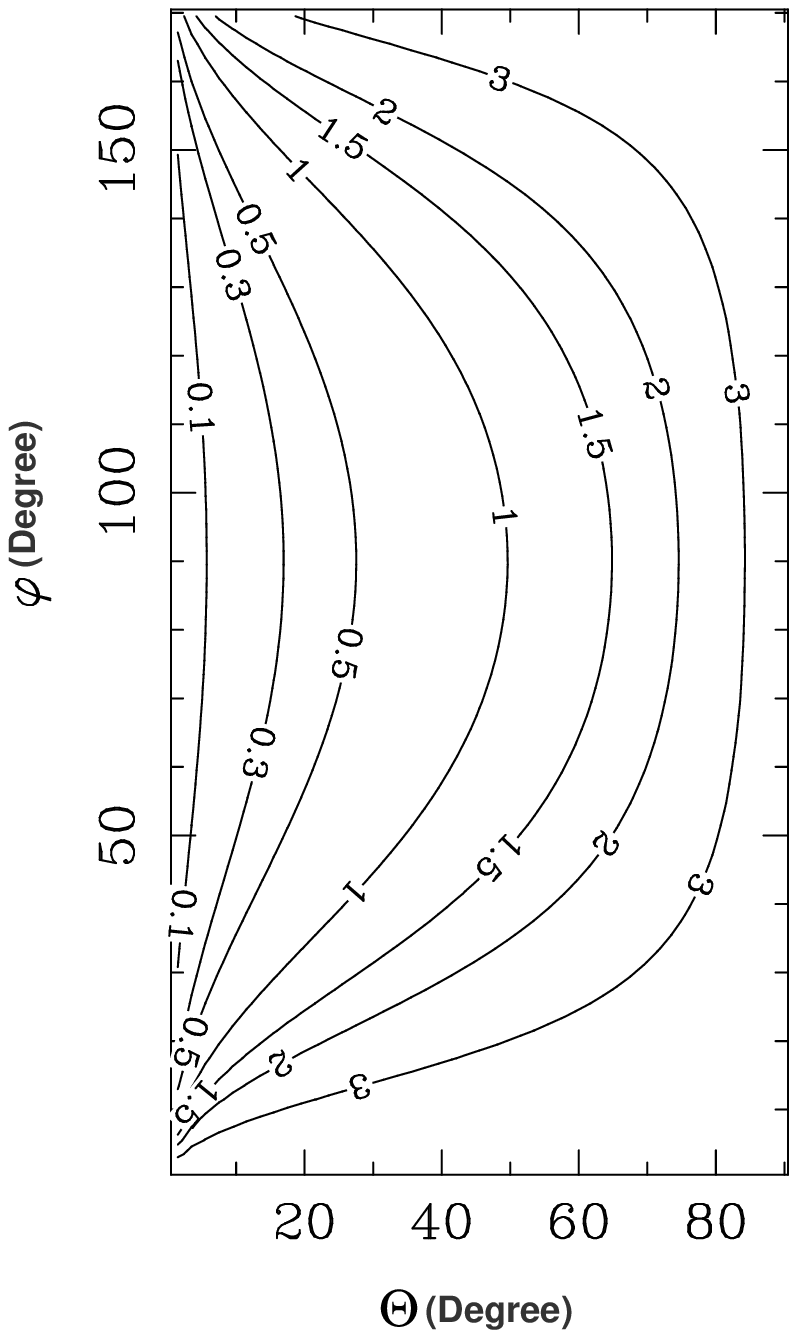} \caption{The geometric function
$g$($\Theta$,$\varphi$).}
\end{figure}

\begin{figure}

\centering \plotone{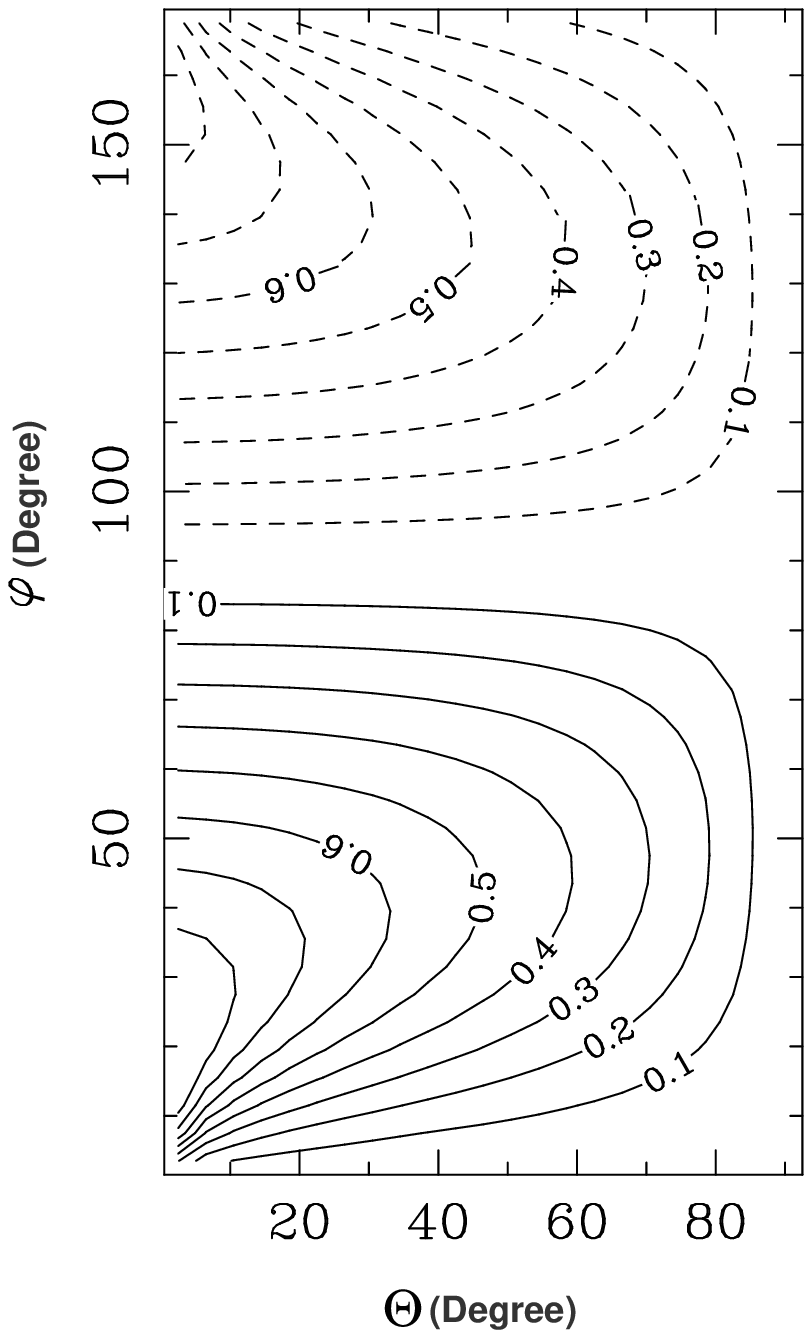} \caption{The geometric function
$f$($\Theta$,$\varphi$).}
\end{figure}

\begin{figure}

\centering \epsscale{1.0} \plotone{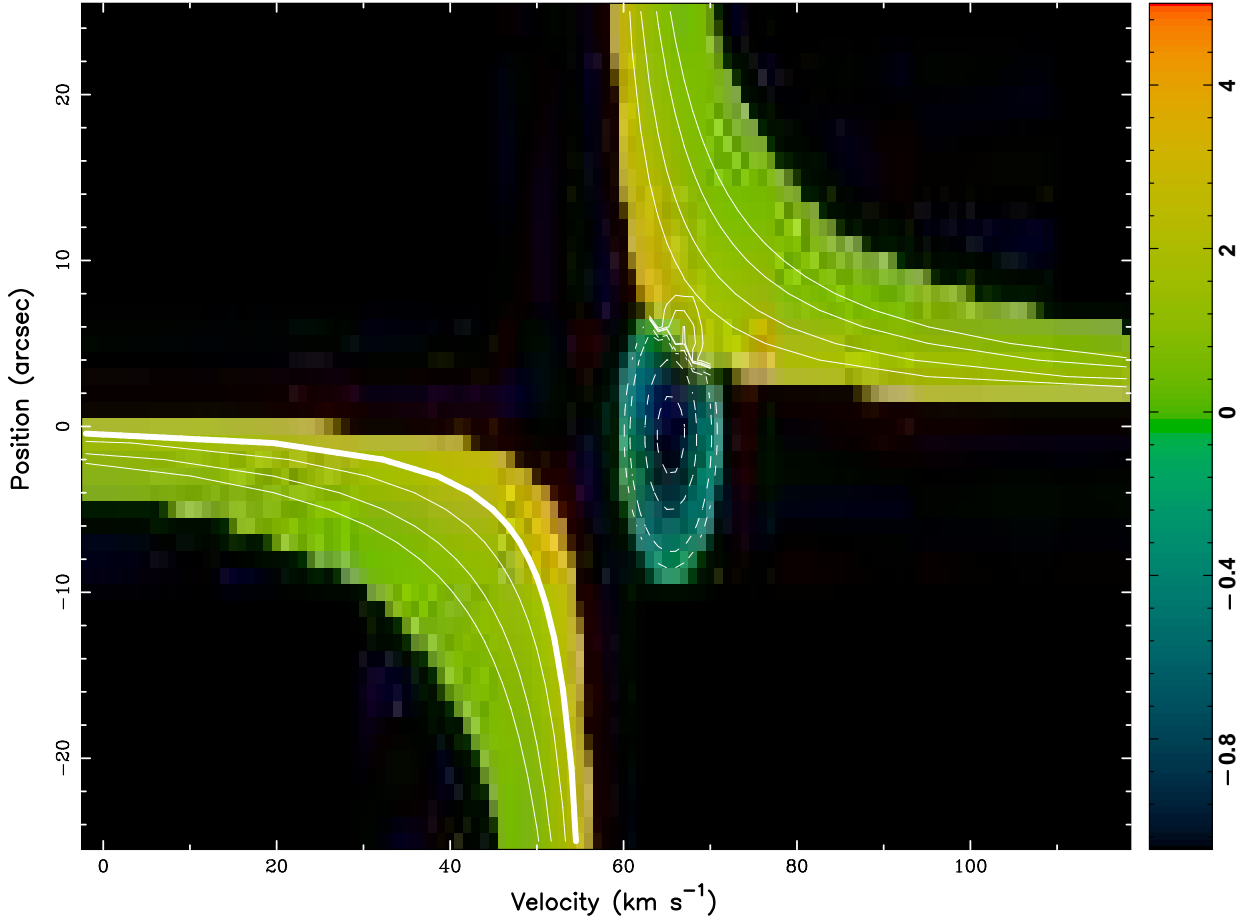} \caption{The PV
diagram calculated from a model incorporating a spherical inflow
with decelerating bipolar outflow for the parameter $V_0 r_{min}
f(\Theta,\varphi)$ in a range between 0.5 and 5.1 km s$^{-1}$ pc
(scaled in color). The position axis (vertical) is a cut along the
major axis of the outflow. The velocity axis (horizontal) is the
radial velocity of the outflow with respect to the LSR. The
systematic velocity of the central object is assumed to be 58 km
s$^{-1}$ and the mean infall velocity of the absorption component
is 8 km s$^{-1}$. The contours correspond to $V_0 r_{min}
f(\Theta,\varphi)$=0.8, 1, 1.4 and 2 km s$^{-1}$ pc. The thick
contour in the blue-shifted outflow corresponds to 2 km s$^{-1}$
pc. The absorption (dashed contours) feature at V=66 km s$^{-1}$
and X=0 arcsec shows the component of spherical inflow. The normalized
absorption intensity is scaled in color.}

\end{figure}

\begin{figure}

\centering \epsscale{0.8} \plotone{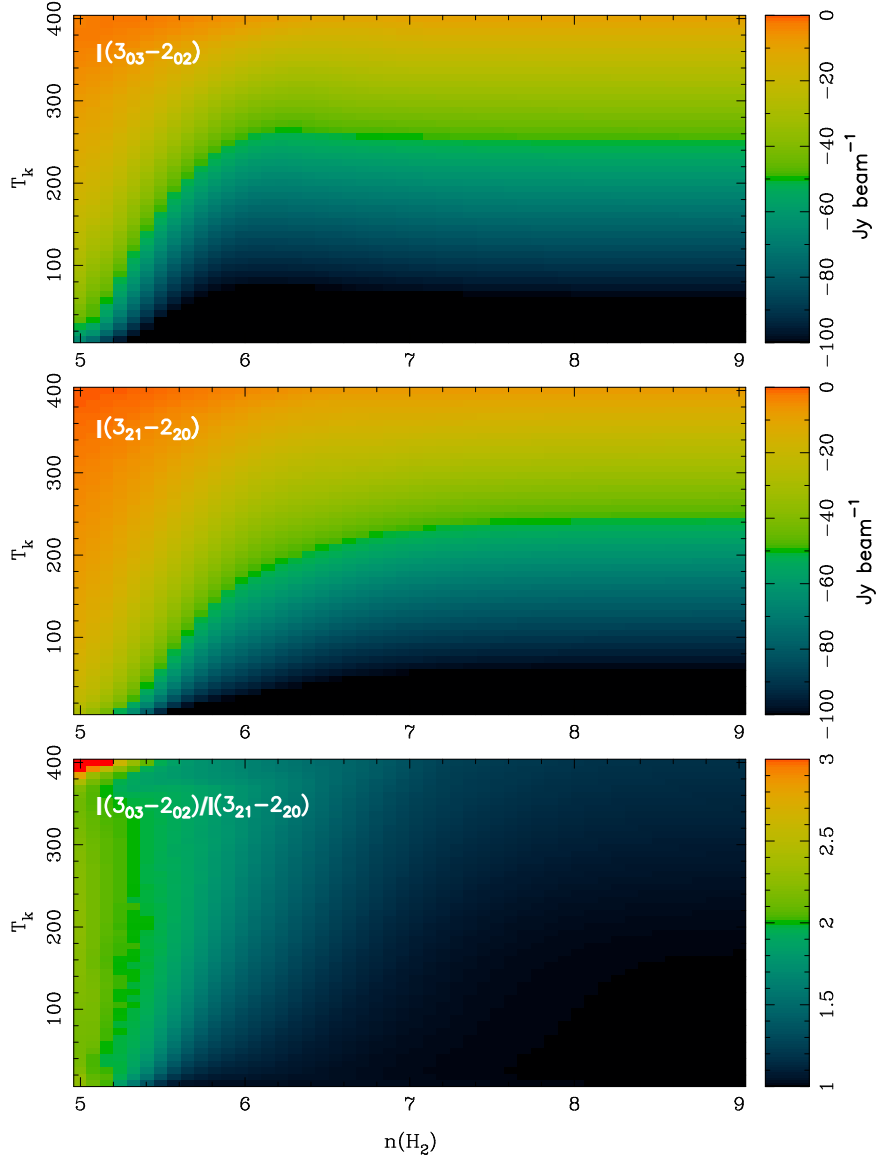} \vspace{-5mm}
\caption{An example shows the LVG analysis results derived with
model C for component K1-3 at 72 km s$^{-1}$. Top panel: The line
intensity of para-H$_2$CO transition I($3_{03}-2_{02}$) as a
function of n$_{\rm H_2}$ and T$_{\rm k}$. Middle panel: The line
intensity of para-H$_2$CO transition I($3_{21}-2_{20}$) as a
function of n$_{\rm H_2}$ and T$_{\rm k}$. Bottom panel: The line
ratio of I($3_{03}-2_{02}$)/I($3_{21}-2_{20}$) as a function of
n$_{\rm H_2}$ and T$_{\rm k}$. }
\end{figure}

\begin{figure}

\epsscale{.80} \plotone{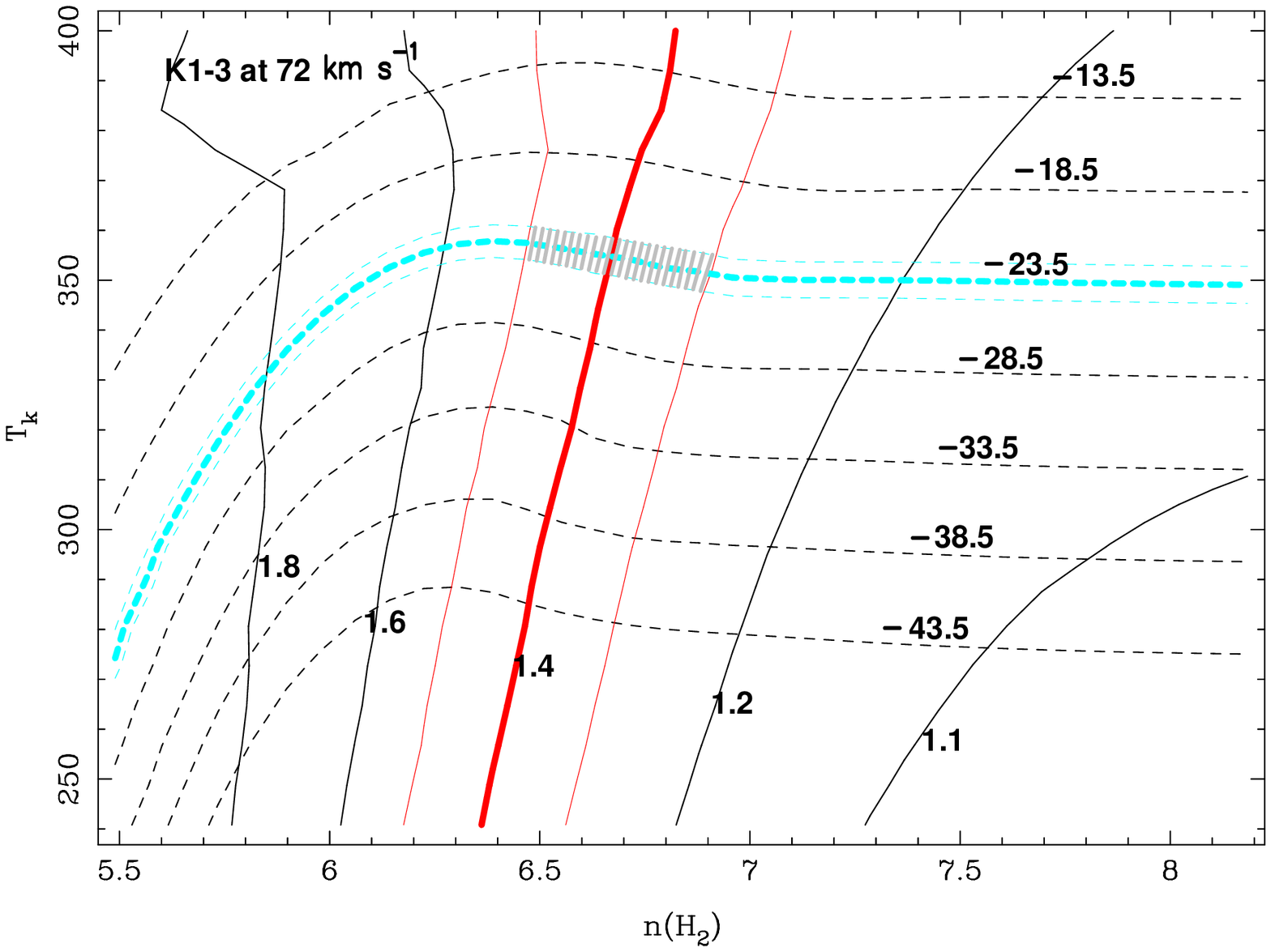} \epsscale{1.0}

\plottwo{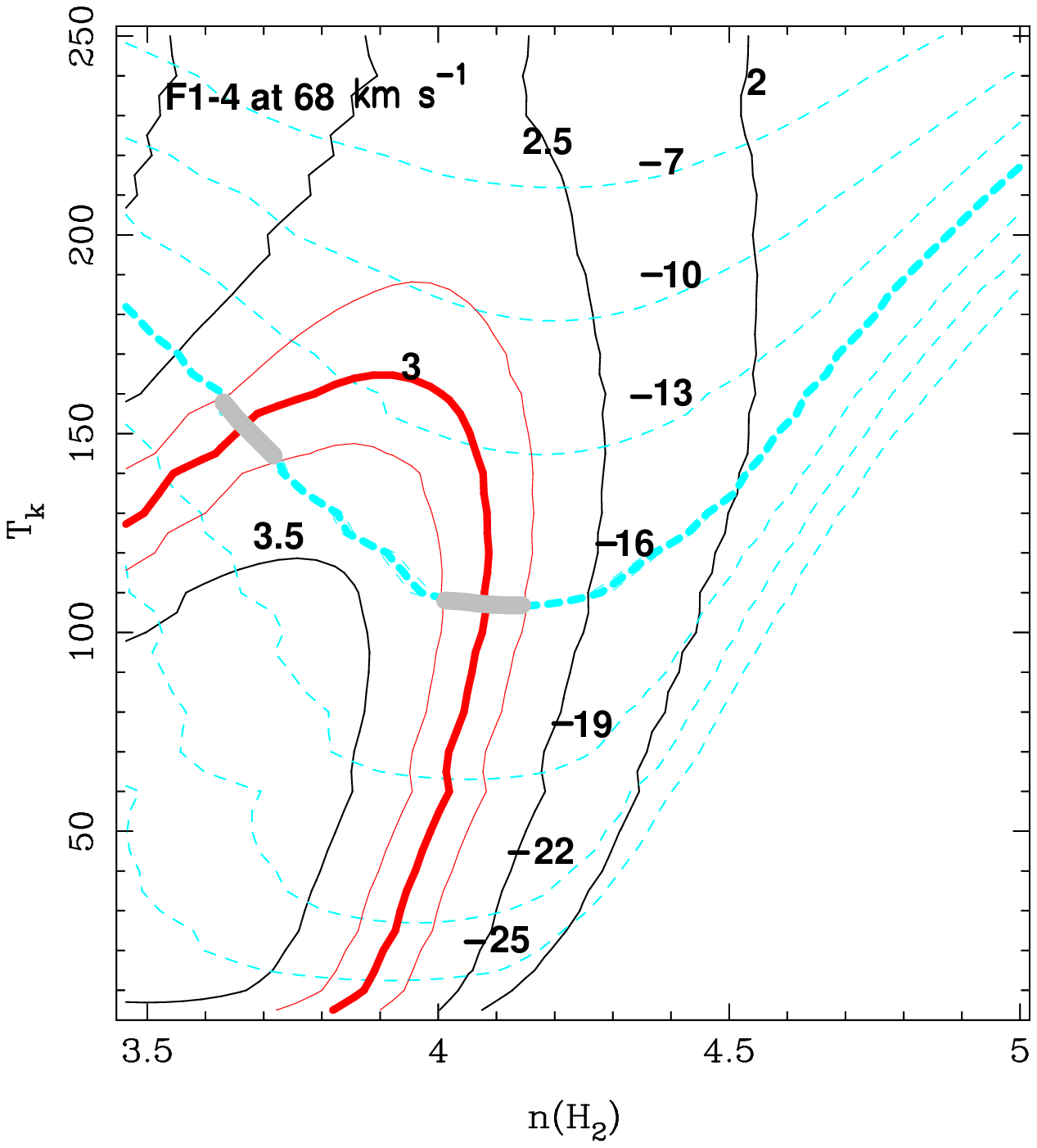}{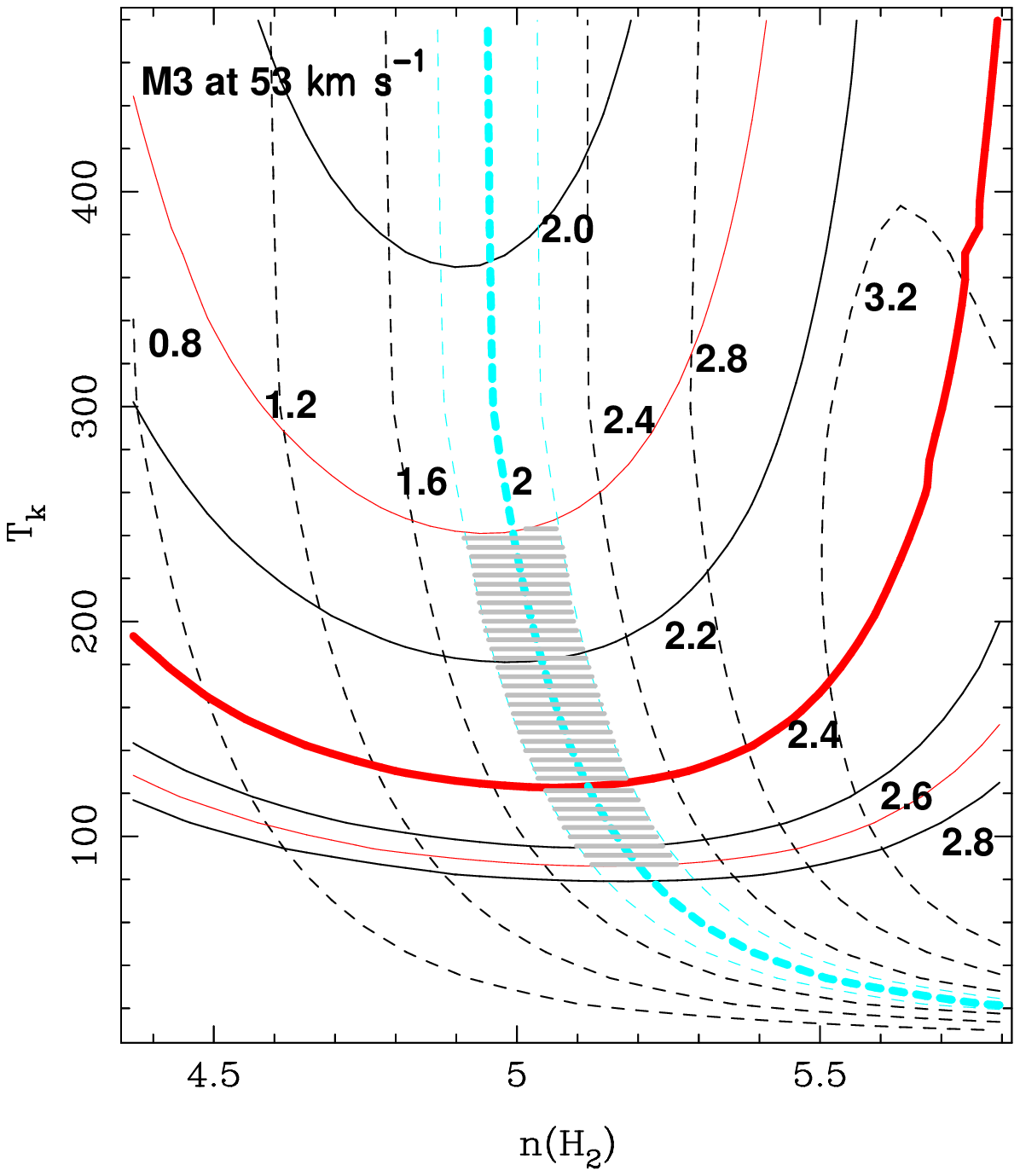} \vspace{-3mm} \caption{Examples of
fitting the observed line intensity and intensity ratio to model C
with the LVG approximation. Dashed lines are the line intensity of
the $3_{03}-2_{02}$ transition. The solid lines are the line ratio
I($3_{03}-2_{02}$)/I($3_{21}-2_{20}$). The thick solid and dashed
lines correspond to the observed values of I($3_{03}-2_{02}$) and
I($3_{03}-2_{02}$)/I($3_{21}-2_{20}$). The hashed grey zones mark
the solution regions in the n$_{\rm H_2}$-T$_{\rm k}$ domain. Top:
for the absorption component K1-3 at 72 km s$^{-1}$. Bottom-left:
for the absorption component F1-4 at 68 km s$^{-1}$. Bottom-right:
for the emission component M3 at 53 km s$^{-1}$. }

\end{figure}


\begin{thebibliography}{}

\bibitem{cB90} Cabrit, S. \& Bertout, C. 1990, ApJ, 348, 530

\bibitem{cv89} Carlstrom, J.E. \& Vogel, S.N. 1989, ApJ, 337, 408

\bibitem{dgg95} de Pree, C.G., Gaume, R.A., Goss, W.M., et al. 1995, ApJ, 451, 284

\bibitem{dgg96} de Pree, C.G., Gaume, R.A., Goss, W.M., et al. 1996, ApJ, 464, 788

\bibitem{dgg98} de Pree, C.G., Goss, W.M. \& Gaume, R.A. 1998, ApJ, 500, 847

\bibitem{dmn00} de Vicente, P., Martin-Pintado, J., Nera, R., et al. 2000, A\&A, 361, 1058

\bibitem{edp75} Evans II, N.J., Davis, J.H. \& Plambeck, R.L. 1979, ApJ, 227, L25

\bibitem{gg90} Gaume, R.A. \& Claussen, M.J. 1990, ApJ, 351,538

\bibitem{gg95} Gaume, R.A., Claussen, M.J., de Pree, C.G., et al. 1995, ApJ, 449, 663

\bibitem{grc04} Goicoechea, J.R., Rodriguze-Fernandez, N.J. \& Cernicharo J. 2004, ApJ, 600, 214

\bibitem{g95} Gordon, M. A. 1995, A\&A, 301, 853

\bibitem{g91} Green, S. 1991, ApJS, 76, 979

\bibitem{h83} Hildebrand, R.H. 1983, QJRAS, 24, 267

\bibitem{klf84}Kahane, C., Lucas, R., Frerking, M.A., et al. 1984, A\&A, 137, 211

\bibitem{ks94} Kuan Yi-Jehng \& Snyder L.E. 1994, ApJS, 94, 651

\bibitem{ks96} Kuan Yi-Jehng \& Snyder L.E. 1996, ApJ, 470, 981

\bibitem{kms96} Kuan Yi-Jehng, Mehringer D. M. \& Snyder L.E. 1996, ApJ, 459, 619

\bibitem{lck91} Lis, D.C., Carlstrom, J.E. \& Keene. J. 1991, ApJ, 380, 429

\bibitem{lgc93} Lis, D.C., Goldsmith, P.F., Carlstrom, J.E., et al. 1993, ApJ, 402, 238

\bibitem{ls99}  Liu, Sheng-Yuan  \& Snyder, L.E. 1999, ApJ, 523, 683

\bibitem{l99}Lizano, S., Heiles, C., Rodriguez, L.F., et al. 1988, ApJ, 328, 763

\bibitem{lu71}Lucy, L. B. 1971, ApJ, 163, 95

\bibitem{mw93}Mangum, J.G. \& Wootten, A. 1993, ApJS, 89, 123

\bibitem{mdw90}Martin-Pintado, J., de Vicente, Wilson, T.L., et al. 1990, A\&A, 236, 193

\bibitem{mpg93}Mehringer, D. M., Palmer, P., Goss, W.M., et al. 1993, ApJ, 412, 684

\bibitem{mgp94} Mehringer, D. M., Goss, W.M. \& Palmer, P. 1994, ApJ, 434, 237

\bibitem{mdp95}Mehringer, D.M., Palmer, P. \& Goss, W. M. 1995, ApJS, 97, 497

\bibitem{mmd95}Miao, Y. T., Mehringer, D. M., Kuan, Yi-Jheng, et al. 1995, ApJ, 445, L59

\bibitem{nbh98} Nummelin, A., Bergman, P., Hjalmarson, A., et al. 1998, ApJS, 117, 427

\bibitem{prg00} Pierce-Price, D., Richer, J.S., Greaves, W.S., et al. 2000, ApJ, 545, L121

\bibitem{rcc93}Raga, A.C., Canto, J., Calvet, N., et al. 1993, A\&A, 276, 539

 \bibitem{rsm88} Reid, M.J., Schneps M.H., Moran J.M. et al. 1988, ApJ, 330, 809

\bibitem{ss74} Scoville, N. Z. \& Solomon, P. M. 1974, ApJ, 187, L71

\bibitem{s77} Shu, F. 1977, ApJ, 214, 488

\bibitem{sbm85} Sutton, E.C., Blake, G.A., Masson, C.R., et al. 1985, ApJS, 58, 341

\bibitem{sjd91} Sutton, E.C., Jaminet, P.A., Danchi, W.C., et al. 1991, ApJS, 77, 255

\bibitem{vgs96} Vacca, W., Garmany, C. \& Shull, J. 1996, ApJ, 460, 914

\bibitem{vgp87} Vogel, S.N., Genzel, R. \& Palmer, P. 1987, ApJ, 316, 243


\end{thebibliography}
\end{document}